\documentclass[aps,prb,showpacs,twoside,twocolumn,10pt]{revtex4-2}
\usepackage[colorlinks=true, citecolor=red, urlcolor=blue ]{hyperref}
\usepackage{times,epsfig,amssymb,amsfonts,amsmath,bm,subfigure,mathtools,amsthm,braket,soul,enumitem,xcolor,physics,graphics,graphicx, float}
\UseRawInputEncoding
\usepackage[normalem]{ulem}
\usepackage{comment, appendix}
\usepackage{epstopdf}

\usepackage{comment}

\begin{document}
\title{
Revealing effects of local dimension on variable-range interacting model \\ by connecting Lieb-Robinson bounds and multipartite entanglement 
}
\author{Keshav Das Agarwal, Debkanta Ghosh, Pritam Halder, Aditi Sen(De)}
\affiliation{Harish-Chandra Research Institute, A CI of Homi Bhabha National Institute,  Chhatnag Road, Jhunsi, Allahabad - 211019, India}

\begin{abstract}

A spin-$s$ variable-range interacting Ising model may display qualitatively different behaviors depending on the fall-off rate of the interactions, as already seen in equilibrium studies of spin-1/2 systems. We propose a dynamical method  using weighted graph states, generated through time evolution that confirms the existence of the transition point in the fall-off rate for the spin-$s$ Ising model. Moreover, the dependence of local dimension on information spreading and multipartite entanglement  profile in this model remains unclear, which we establish here. In particular, our analysis shows that 
the maximum of genuine multipartite entanglement (GME) with the fall-off rate serves as a clear indicator of the information spreading, which aligns with changes in the profile of the Lieb-Robinson bound. Further, in the case of an open chain, the spread of information is related to the divergence in the first derivative of GME. Additionally, we validate this signature by performing a scaling analysis of the time-averaged mutual information.

\end{abstract}

\maketitle

\section{Introduction}

The overwhelming focus of theoretical breakthroughs in quantum technologies involves mostly two-level quantum systems (qubits), although the majority of physical systems contain several levels that can be approximated as qubit configuration \cite{nielsenchuang}. Moreover, multilevel quantum systems known as qudits can exhibit a non-trivial entanglement structure \cite{horodecki_pla97,bnd_ent_horodecki_prl98,craft_huber_guhne_prl_2018}, which can also be confirmed by their violation of  Bell inequalities \cite{cjlmp_2002,son_lee_kim_prl2006,Dada2011}.  They are also demonstrated to provide superior performance compared to their two-level counterparts in a myriad of quantum information processing tasks, including  quantum secure key rates \cite{bechmann2000,cerf_qkd_prl_2002,nikolopoulos2006,sheridan2010,Sasaki2014,Bouchard2018,cozzolino2019,araujo2023,bulla2023}, quantum simulation and computation \cite{Neeley2009,Kaltenbaek2010}, quantum telecloning \cite{nagali2010,Bouchard2017}, the robust test of quantum steering \cite{marciniak2015,Srivastav2022}, dense coding protocol against noise \cite{patra2023dimensional}, and quantum thermal machines \cite{ correra2014,wang2015,santos2019,Dou2020,usui2021, ghosh2022, konar2023}. Additionally, quantum spin models such as nearest-neighbor and next-nearest-neighbor Heisenberg models  with spin-$1$ and spin-$3/2$, bilinear-biquadratic spin-$s$ models \cite{Lai1974, sutherland1975,  Takhtajan1982, Babujian1982, AKLT_prl87, fath1991, fath1993, buchta2005, mao2021} may display rich phase diagrams \cite{spin1_ising_71,sivardier1972, AKLT_nnn_prl96} which cannot be found with spin-$1/2$ chains.
Along with theoretical developments, experimental evidence shows that qudit-based quantum computers \cite{Zhou_2003,Hosten_2005,hall2005cluster, joo_2007,Zhou_2009,Raussen_2017, Booth_2023,Imany2019, Wang2020,Sawant_2020,expt2020,Hrmo2023} are more efficient than qubit-based ones due to the larger computing space of each particle and improved error correction codes \cite{chau97,braun98,hu08,Grassl2018,gunder2020,schmidt2022,gunder2022b}.

Cluster states \cite{Briegel2001} are resource for measurement-based quantum computation (MBQC) \cite{Raussendorf2003, Briegel2009, spt_mbqc}, and are also studied as a ground state of cluster Hamiltonians obtained by gauging nearest-neighbor (NN) Hamiltonians \cite{nSPT_Gclu} having symmetry-protected topological order \cite{Son2012}. The energy spectrum of cluster Hamiltonian is degenerate in the case of an open boundary condition, which can be used to form edge states \cite{nSPT_Gclu}. Interestingly, they can also be prepared as a dynamical state of the NN Hamiltonian \cite{Briegel2001, Raussendorf2003, Briegel2009}. However, in recent years, there has been a significant surge in interest to extend the studies of many-body systems with short-range (SR) interactions to those with variable-range (VR) ones, inevitably appearing in physical substrates such as trapped ions \cite{HAFFNER2008,iontrap08,iontrap09,iontrap12,Islam_2011,Britton_2012,iontrap_cirac04,expt2020}, Rydberg atoms \cite{Choi2016,Rispoli2019}, ultracold atoms in optical lattices \cite{Mandel2003, Cramer2013, Landig2016,Gross2017,camacho2107}, along with superconducting qubits \cite{PRXQuantumsuper,ying23}. They are characterized by the interaction between a pair of particles, $(x,y)$, residing at a distance $r=|x-y|$, which falls off as $g_{xy} \propto 1/r^{\alpha}$ with \(\alpha>0\), representing  the range of interactions \cite{sachdev_2011,Campa2014,Maskara2022,defenu_rmp2023}. In both equilibrium and out-of-equilibrium scenarios, by varying the fall-off parameter $\alpha$, unique observations including dynamical quantum phase transition \cite{Zhang2017,defenu2019,halimeh2020,lakkaraju2023framework, Defenu2024}, metastable phases \cite{Defenu2021, Giachetti2023}, and information scrambling \cite{chen2019,tran2020,kuwahara2020} are reported, which are otherwise absent in SR interacting systems. Interestingly, long-range (LR) interactions have been demonstrated to have a positive impact on the generation of highly mutlipartite entangled states \cite{anuradha2023production,ghosh2023entanglement}, quantum heat engine \cite{Solfanelli2023}, sensing \cite{Yousefjani2023,monika2023better}, state transfer \cite{eldredge2017, tran2021} and computation \cite{lewis2021,Lewis2023,ghosh2025measurementbasedquantumcomputationvariablerange}. Note further that the long-range Ising interaction generalizes the cluster state to the weighted graph state (WGS) with both periodic (PBCs) and open boundary conditions (OBCs) \cite{dur2005, hein2006entanglement, anil2022, ghosh2023entanglement}.

The propagation of information in such models varies with the range of interaction. In particular, the Lieb-Robinson bound (LRB) establishes that, for a time evolution governed by any local Hamiltonian, information propagates at a finite velocity, denoted as \( v_{\text{LR}} \). This results in the formation of a causal cone in space-time, beyond which correlations decay exponentially with distance. Mathematically, this behavior is expressed as  $\lVert [A_{\vec{0}}(t), B_{\vec{r}}] \rVert \leq c_{\text{LR}} e^{-\frac{|\vec{r}| - v_{\text{LR}}t}{\xi_{\text{LR}}}}$ for local Hamiltonians, where \( c_{\text{LR}} \), \( \xi_{\text{LR}} \), and \( v_{\text{LR}} \) are parameters that depend on the specific details of the Hamiltonian, the underlying lattice structure, and the choice of operators \( A \) and \( B \) \cite{LRB_og, Hastings2006, Nachtergaele2006, Cheneau2012}. Here, \( A_{\vec{0}} \) is defined at the origin, while \( B_{\vec{r}} \) is located at a distance \( \vec{r} \). This gives a causal structure to quantum effects of local interactions even in the nonrelativistic scenarios. The study of LRBs in LR systems shows sublinear light cones, and $v_{\text{LR}}$ can increase with time, showing nonlocal properties \cite{Eisert2013,Gong2014,Foss-Feig2015,Tran2019,chen2019,Else2020,Yin2020,kuwahara2020,tran2020,Tran2021_LRB,Chen2021,Gong2023,lemm2024EnLRB}.

We propose and establish that the WGS framework can be explored to study the effect of LR interactions along with local spin dimension through the dynamics of mutual information (MI), the information propagation quantified by LRB, and the genuine multipartite entanglement (GME). Specifically, the profile of the time-averaged MI of the WGS, between sites with short distances exhibits a transition in its functional behavior at fall-off rates which depend logarithmically on the dimension of the local Hilbert space. We exhibit that although the LRBs depict the nonlocal spread of information at a particular fall-off rate, independent of both the local spin dimension $d$ and the spatial dimension $D$,  the maximum information spread does display a logarithmic dependence of $d$ on the fall-off rates. Further, in the case of OBC, our investigation reveals that the GME of the WGS becomes nonanalytic at times according to the relation given by the LRBs. Our analysis also demonstrates the existence of dynamically nonlocal behavior for the spin-$s$ VR Ising model that is already known for spin-\(1/2\) systems.

The paper is organized in the following manner. In Sec. \ref{sec:spin-d wgs}, we briefly introduce the structure of the higher-dimensional WGS generated by a spin-$s$ long-range Ising Hamiltonian and the form of reduced density matrices. The influence of local spin dimension on mutual information and LRB is presented in Sec.~\ref{sec:mutual inf} and \ref{sec:LRB}, respectively. We exhibit a correspondence between LRB and the nonanalytic behavior of genuine multipartite entanglement in Sec.~\ref{sec:GGM} before concluding in Sec. \ref{sec:conclusion}.

\section{Generation of multiqudit weighted graph state}
\label{sec:spin-d wgs}
% {\it Generation of multiqudit weighted graph state.} 

A graph $G=(V,E)$ can be used to describe a lattice of $N$ spin-$s$ systems interacting via long-range interactions, with a vertex set $V=\{1,2,\ldots,N\}$ of $d=2s+1$ level quantum systems and $E\subseteq [V]^2$ being the weighted edges. Each edge $E_{xy}\in E$ denotes the interaction strength between qudits, $x,y\in V$, at positions $\vec{x}$ and $\vec{y}$ respectively, of the underlying lattice. For long-range systems interacting via Coulomb-like potential, we have  $E_{xy}=1/|\vec{r}_{xy}|^\alpha$ with $\vec{r}_{xy}=\vec{y}-\vec{x}$ and fall-off rate $\alpha$ keeping $|\vec{r}_{xy}|\geq1$ ($|\vec{r}_{xy}|=1$ corresponds to nearest neighbors). The qudit weighted graph state is generated by initializing each vertex in the state $\ket{+}_d=\frac{1}{\sqrt{d}}\sum_{m=0}^{d-1}\ket{m}$ and evolving each pair of qudits $(x,y)$ via spin-$s$ Ising interaction $H^d_{xy}=\sum_{a_x,a_y=0}^{d-1} a_x a_y \ket{a_x a_y}\bra{a_x a_y}$ \cite{Zhou_2003,dur2005,hein2006entanglement,ghosh2023entanglement,anil2022}. Therefore, the total Hamiltonian, and the corresponding WGS of system size $N$ are given by 
\begin{eqnarray}
H^d(\alpha) &=& \sum_{\mathclap{x<y\in V}} H^d_{xy}/|\vec{r}_{xy}|^\alpha\ ;\nonumber\\
\ket{\psi(\alpha,t)} &=& \frac{1}{d^{\frac{N}{2}}}\sum_{\mathclap{\vec{a}}}e^{-i\mathcal{E}_{\vec{a}}t}\ket{\vec{a}}
\end{eqnarray}
respectively, where $\mathcal{E}_{\vec{a}}(\alpha)=\sum_{x<y\in V}a_x a_y/|\vec{r}_{xy}|^\alpha$ and $\ket{\vec{a}}\equiv\ket{a_1, a_2,\dots, a_N}$ with $a_{x,y}\in \{0,1,\dots,d-1\}$ forming the $d^N$ computational (eigen)basis. The qudit cluster states and their uses in MBQC have been studied \cite{Zhou_2003}, where the evolving Hamiltonian involves only NN interactions, i.e., $H^d(\alpha\to\infty)$. However, the multiqudit WGS with variable-range interaction and its entanglement properties still remain largely unexplored. To achieve the goal, we compute reduced density matrices of a subsystem, following the generalized PEPS formalism \cite{Hartmann2007, Brell2015}. For example, the single-site reduced density matrix for the first site takes the form
\begin{equation}
    [\rho_{\{1\}}]_{\mu\nu}=\frac{1}{d}\prod_{q=2}^N\bigg(\frac{1}{d}\sum_{k=0}^{d-1}\exp[i(\mu-\nu)k \:\frac{t}{|\vec{r}_{1q}|^\alpha}]\bigg),
\end{equation}
with $\mu,\nu\in\{0, 1,\dots,d-1\}$. In a similar fashion, it is possible to obtain all the reduced density matrices of the dynamical state after making an arbitrary partition in the system (see Appendix~\ref{app:reduce_density_matrix}). 

\section{Impact of local spin dimension in information propagation}
\label{sec:impact}

Tuning the fall-off rate $\alpha$ allows the system Hamiltonian to interpolate between nearest-neighbor and long-range regimes. Consequently, the characteristics of information transfer across the lattice depend sensitively on the decay exponent governing the interaction strength, which has been explored within the framework of LRB~\cite{Eisert2013,Gong2014,Foss-Feig2015,Tran2019,chen2019,Else2020,Yin2020,kuwahara2020,tran2020,Tran2021_LRB,Chen2021,Gong2023,lemm2024EnLRB}. On the other hand, the effect of the local Hilbert space on it has yet to be systematically investigated. Here, we address this question by analyzing the scaling of the time-averaged mutual information~\cite{nielsenchuang, Cerf_1997, Cover_2005, Groisman_2005, Chisholm_2024} and subsequently we show how LRBs dynamically encode the dependence on local dimensionality through maximal information transfer.
% {\it Identifying the effect of local spin dimension in mutual information. }

\subsection{Propagation of mutual information}
\label{sec:mutual inf}

\begin{figure*}
    \centering
    \vspace{-3.0mm}
    \includegraphics[width=0.75\linewidth]{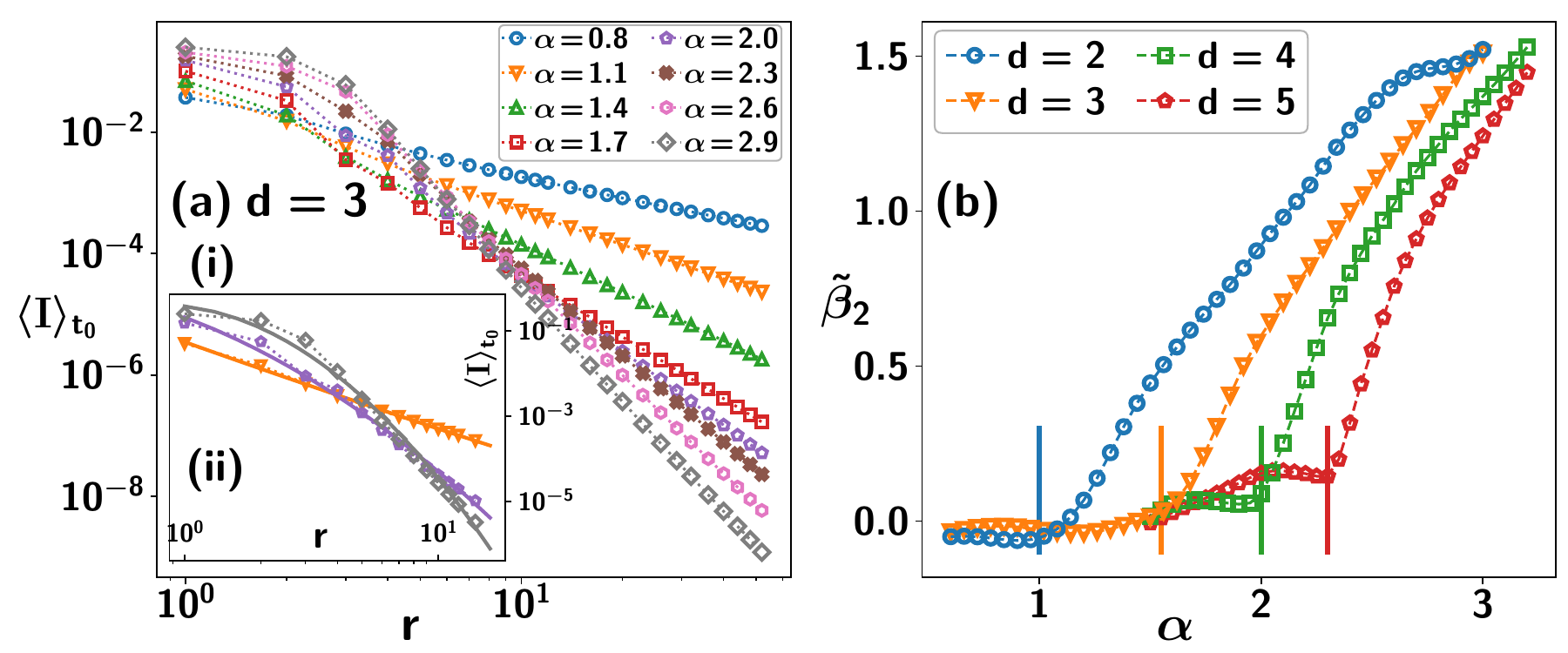}
    \vspace{-3.0mm}
    \caption{(a) Time-averaged mutual information, $\langle I\rangle_{t_0}(\alpha, r)$(ordinate) for qutrit (spin-$1$, $d=3$) WGS in $1$D  with OBC, against the distance $r$ (abscissa) between two sites, in logarithmic scales, for various fall-off rates $\alpha$. The average is taken from $0$ to $t_0=15\pi$. Although, $\log\langle I\rangle_{t_0}$ decreases linearly with $\log r$ at large $r$ for all $\alpha$, its behavior changes from linear to quadratic on increasing $\alpha$ at small $r$. The quadratic fitting is done from $r=1$ to $15$ which shows good agreement in inset (a)(ii). (b) Quadratic coefficient $\tilde{\beta}_2$ (ordinate) of  $\log\langle I\rangle_{t_0}(\alpha, r)$ scaling with $\log r$,  against fall-off rate $\alpha$ (absissca), showing $\alpha\sim\log_2d$ (vertical bars) in the change of $\langle I\rangle_{t_0}(\alpha, r)$ profile from linear to quadratic (in log scales). Here, $N=4000$, and all the axes are dimensionless.} 
    \label{fig:alpha_mi}
 \end{figure*}

Considering a one-dimensional ($1$D) lattice consisting of $N$ sites with OBC and LR interactions, we compute quantum mutual information between two spatially separated subsystems, $A$ and $B$, $I(A:B)\equiv I(\alpha,r,t)$ with the subsystems $A=\{1\}$ and $B=\{r+1\}$ with $r(< N)$. Here, $I(A:B)=S(\rho_A)+S(\rho_B)-S(\rho_{AB})$~\cite{nielsenchuang, Cerf_1997, Cover_2005, Groisman_2005} serves as a measure of total correlation, both classical and quantum, where $S(\rho_A)=-\Tr[\rho_A\log_2\rho _A]$ is the von Neumann entropy of the reduced subsystem, $\rho_A=\Tr_B[\rho_{AB}]$. We observe that depending on the value of $\alpha$, it collapses and revives with time (see Appendix~\ref{app:mutual_info}). In particular, for small values of $\alpha$, the site at $A$ interacts with the sites at a distance $r$ with a higher strength compared to that of high $\alpha$. Therefore, due to monogamy \cite{qmi_monogamy}, $I(\alpha, r, t)$ for small $\alpha$ and $r$ (sites near $A$) immediately reaches maximum and then decreases to zero with only small oscillations afterwards, i.e., its value is nonvanishing only for a short time, while for large $\alpha$, relative interactions at small $r$ are prominent, and hence $I(\alpha, r, t)$ remains nonvanishing at larger times as well. Therefore, the finite time-average of MI, defined as $\langle I \rangle_{t_0}\equiv\langle I \rangle_{t_0}(\alpha,r)=\frac{1}{t_0}\int_0^{t_0} I(\alpha,r,t)dt$ must be positively correlated with $\alpha$ at small $r$, while it should decrease with increasing $\alpha$ at larger distances. We illustrate this behavior in Fig.~\ref{fig:alpha_mi} for $d=3$ and $t_0=15\pi$, with $\langle I \rangle_{t_0}$ having a power law scaling, i.e., $\log \langle I \rangle_{t_0} \sim -\tilde{\beta}_1(\alpha) \log r$, in the case of a large distance $(r\gtrsim10)$. 
Here, $t_0$ can be taken to be any value larger than \(2\pi\), i.e., taking $ t_0>> 2\pi$  does not change any qualitative behavior of \(\langle I \rangle_{t_0}\) . For small separation distance, i.e., $r\lesssim10$, $\log \langle I \rangle_{t_0}$ shows different scaling with $\log r$, i.e., $\log \langle I \rangle_{t_0}\sim -\tilde{\beta}_1(\alpha)\log r$ (linear) for small $\alpha$ ($\alpha\ll 1$) while $\log \langle I \rangle_{t_0}\sim -\tilde{\beta}_2(\alpha)(\log r)^2$ (quadratic)  for $\alpha\gg 1$ (see Fig.~\ref{fig:alpha_mi}). When $r\lesssim 10$ [Fig.~\ref{fig:alpha_mi}(a)(ii)], the behavior of time-averaged MI can be expressed as 
\begin{eqnarray}
 \log \langle I \rangle_{t_0} = -\tilde{\beta}_2(\alpha)(\log r)^2-\tilde{\beta}_1(\alpha)\log r +\tilde{\beta}_0(\alpha),
\end{eqnarray}
where $\tilde{\beta}_2$ captures the logarithmic dependence on $\alpha$ of a distinct change in the spread of correlations as shown in Fig.~\ref{fig:alpha_mi}(b). This implies   $\log \langle I \rangle_{t_0}$ at small separation changes its behavior from linear to quadratic at $\alpha\approx \log_2d$ which demonstrates the logarithmic dependence on local Hilbert space dimension on the fall-off rate.  
\begin{figure*}
    \centering
    \vspace{-3.0mm}
    \includegraphics[width=0.65\linewidth]{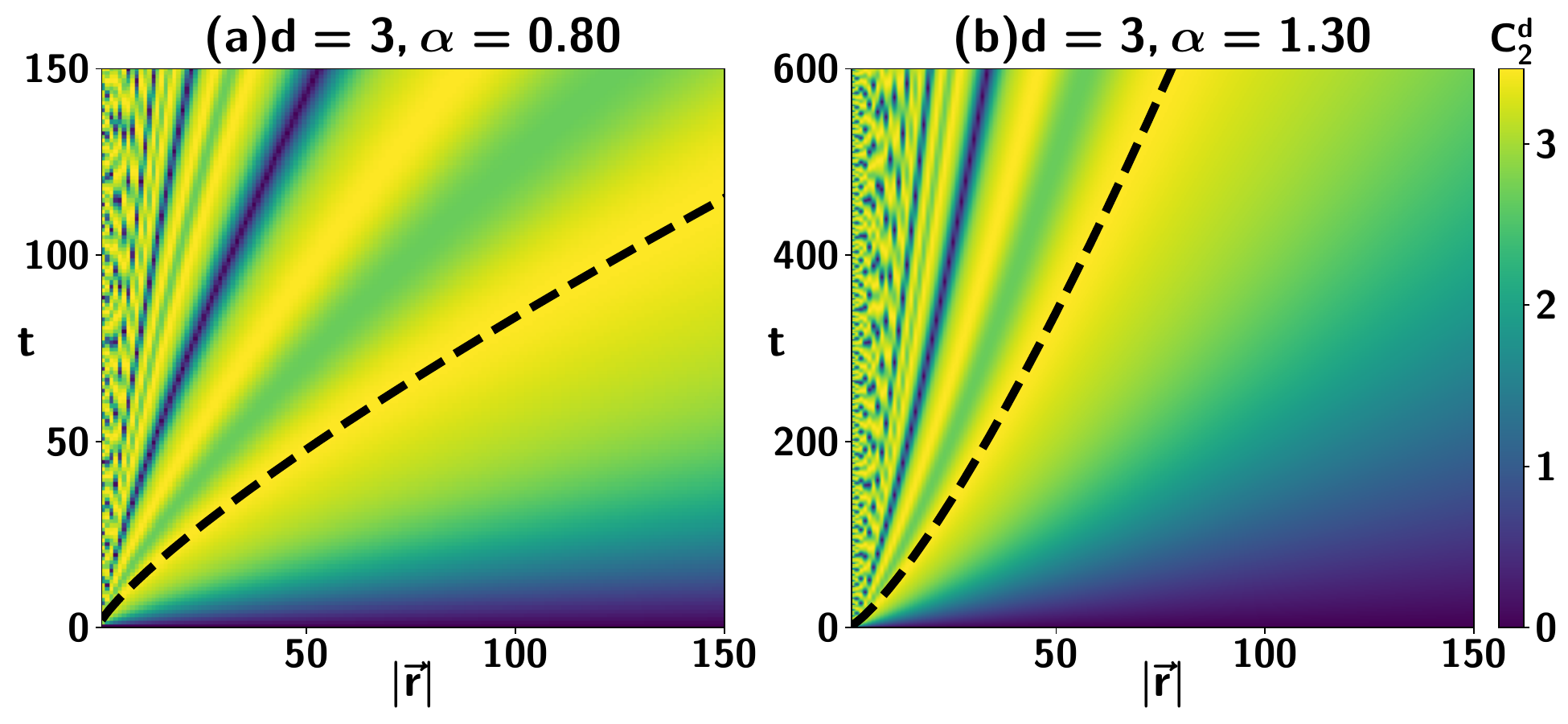}
    \vspace{-3.0mm}
    \caption{Information spread for $d=3$ in the LR Ising model on general lattice, quantified by the commutator $C^d_2$ (colorbar) [Eq.~(\ref{eq:c2})] for fall-off rates (a) $\alpha=0.8$ (b) $\alpha=1.3$ in the time $t$ (ordinate)--distance $|\vec{r}|$ (abscissa) plane. The dashed (black) line represents the $t_{|\vec{r}|}d=2\pi|\vec{r}|^\alpha$ line [Eq.~(\ref{eq:lrb_main})] showing maximum $C^d_2$. All the axes are dimensionless.}
    \label{fig:lrb_d3}
 \end{figure*}

\subsection{Maximal information spreading via LRB} 
\label{sec:LRB}
Lieb-Robinson bounds capture the information (or error) propagation in the system \cite{Eisert2013,Gong2014,Foss-Feig2015,Tran2019,chen2019,Else2020,Yin2020,kuwahara2020,tran2020,Tran2021_LRB,Chen2021,Gong2023,lemm2024EnLRB} and can be used to identify the dynamically nonlocal or local effects in the dynamics of the LR quantum systems. For the \mbox{spin-$s$} LR Ising model, on any lattice of spatial dimensions $D$, the maximum allowed spreading is quantified by the maximum of the Schatten $p$-norm of the commutator, $\lVert [A_{\vec{0}}(\alpha,t), B_{\vec{r}}] \rVert_p = C^d_p(\alpha, \vec{r}, t)$, where
$A_{\vec{0}}(\alpha,t) = e^{-iH_d(\alpha)t}A_{\vec{0}}e^{iH_d(\alpha)t}$ is the time evolution of $A_{\vec 0}$ in the Heisenberg picture. Now, $C^d_p(\alpha,\vec{r}, t)$ can be computed from the $d^2\times d^2$ matrix $\bar{C}(d, \alpha, \vec{r},  t)$ (see Appendix~\ref{app:lrb}), with the entries
\begin{equation}
\bar{C}^{a_{\vec{0}},a_{\vec{r}}}_{b_{\vec{0}},b_{\vec{r}}} = \left(e^{ib_{\vec{r}}(b_{\vec{r}}-a_{\vec{0}})\frac{t}{|\vec{r}|^\alpha}} - e^{ia_{\vec{r}}(b_{\vec{0}}-a_{\vec{r}})\frac{t}{| \vec{r}|^\alpha}}\right)A^{a_{\vec{0}}}_{b_{\vec{0}}} B^{a_{\vec{r}}}_{b_{\vec{r}}},
\end{equation}
where $X^{a_{\vec{k}}}_{b_{\vec{k}}} = \langle b_{\vec{k}}|X_{\vec{k}}|a_{\vec{k}} \rangle$ (\(X=A, B, \bar{C}\)) and all $a,b\in\{0, 1,\dots, d-1\}$. Note that this makes the analysis independent of the lattice structure (and the boundary conditions), with lattice structure encoded in a set of values of $|\vec{r}|\geq 1$. We use the Frobenius (Hilbert-Schmidt) norm $p=2$ (see Appendix~\ref{app:lrb} for the $p=\infty$ case), which, for an arbitrary operator, $M:\mathbb{C}^{d^N}\to\mathbb{C}^{d^N}$, is given by $\lVert M \rVert_2 = \sqrt{\frac{1}{{d^N}}\text{Tr}(MM^\dagger)}$ and is a self-dual norm.
The prefactor $\frac{1}{{d^N}}$ eliminates the dependence on local Hilbert space dimension. Mathematically, for the given Hamiltonian $H_d(\alpha)$, we can write 
\begin{equation}
    C^d_2(\alpha,\vec{r}, t) \!=\! \frac{2}{d}\!\sqrt{\sum_{\mathclap{\quad\substack{a_{\vec{0}}, a_{\vec{r}}=0\\ b_{\vec{0}}, b_{\vec{r}} = 0}}}^{d-1}\sin^2\!\!\left(\!\frac{(a_{\vec{0}}\!-\!b_{\vec{0}})(a_{\vec{r}}\!-\!b_{\vec{r}}) t}{2|\vec{r}|^\alpha}\! \right) \!\!| A^{a_{\vec{0}}}_{b_{\vec{0}}} B^{a_{\vec{r}}}_{b_{\vec{r}}}|^2}.
    \label{eq:c2}
\end{equation}
% From Eq.~\eqref{eq:c2} it is clear that 
For arbitrary $\vec{r}$ and $t$, the maximization of $C_2^d$ over the set of qudit operators with the fixed norm  $||A||=||B||=d$ is attained when $A_{\vec{0}}( B_{\vec{r}})=(d|+\rangle_d\langle+|)_{\vec{0}(\vec{r})}$ which connects the LRBs to the WGS formalism. The locus of the maximum value of $C_2^d$, denoted by ${C^d_2}^{\max}$ in the $(|\vec{r}|, t)$-plane, defines the light cones of the dynamics (see Fig. \ref{fig:lrb_d3}).
For a multiqubit scenario, $C^2_2(\alpha,\vec{r}, t) = 2| \sin(t/2|\vec{r}|^\alpha)|$ where the light cone is defined by the time $t_{|\vec{r}|}=\pi|\vec{r}|^\alpha$, above which the site $|\vec{r}|$ stays inside the light cone and at $t_{|\vec{r}|}$, ${C^2_2}^{\max}=2$ is attained at the earliest for any given site $\vec{r}$.
Outside the light cone, i.e., for long distances $|\vec{r^\prime}|\gg (t_{|\vec r|}/\pi)^{\alpha^{-1}}$ and short times, the correlation  $C^2_2(\alpha,\vec{r^\prime},t_{|\vec{r}|})\sim t_{|\vec r|}/|\vec{r'}|^\alpha$ decays according to a power-law.

For $d>2$, we numerically analyze the behavior of $C_2^d(\alpha,\vec r, t)$. Specifically, Fig.~\ref{fig:lrb_d3} illustrates the qutrit scenario in  which the first maximum ${C_2^{3}}^{\max}\approx 3.47$ is attained at $t_{|\vec r|}=2\pi|\vec r|^\alpha/3$, thereby characterizing the light cone. Similarly, numerical investigations confirm that for arbitrary local spin dimension, the light cone is determined by $t_{|\vec r|}=2\pi|\vec r|^\alpha/d$ (see Appendix~\ref{app:lrb} for $d=4$). For $|\vec r|=1$, i.e., for the NN site, $t_{|\vec r|=1}=2\pi/d$ regardless of the fall-off rate $\alpha$, at which the qudit cluster state is generated by the spin-$s$ Ising model. In general, for $|\vec r|\geq 1$, we have 
\begin{equation}
    %\vspace{-8.0mm}
    \alpha=\log_{|\vec{r}|}d +\log_{|\vec{r}|} (t_{|\vec{r}|}/2\pi),
    \label{eq:lrb_main}
\end{equation}
which establishes the explicit logarithmic dependence of $d$ over $\alpha$ as also seen in the case of $\log\langle I \rangle_{t_0}$. In the region outside the light cone, specifically at a given time $t$, in the regime $|\vec {r}|\gg (td/2\pi)^{\frac{1}{\alpha}}$, the correlation $C^d_2(\alpha,\vec{r},t)\sim (d-1)d(d+1)t/|\vec{r}|^\alpha$, i.e., it decays polynomially with distance. Within the light cone, subsequent ${C_2^d}^{\max}$ is achieved at $t_{n,|\vec{r}|}=2\pi n|\vec{r}|^\alpha/d$ ($n\in\mathbb{N}$), when $n$ and $d$ are coprimes, while $C_2^d(\alpha, |\vec{r}|, t)$ vanishes at $t_{m,|\vec{r}|}=2\pi m|\vec{r}|^\alpha$ ($m\in\mathbb{N}$). This results in a nested structure of global maxima and minima of $C_2^d(\alpha, |\vec{r}|, t)$, as evident from Fig.~\ref{fig:lrb_d3} (see Appendix~\ref{app:lrb} for $d=4$). Denoting $v_\alpha(t)=\frac{\mathrm{d}|\vec{r}|}{\mathrm{d}t_{|\vec{r}|}}$ as the speed at which the first envelope of ${C^d_2}^{\max}$ moves, we obtain $v_\alpha(t)\propto t^{\frac{1}{\alpha}-1}$, i.e., $v_{\alpha<1}(t)$ increases with time, showing dynamically nonlocal behavior for $\alpha<1$. When $\alpha>1$, the velocity $v_{\alpha>1}(t)$ decreases polynomially with time whereas the velocity $v_{\alpha=1}$ is constant for strictly linear light cone at $\alpha=1$. Therefore, the velocity \( v_{\alpha}(t) \) increases with time in the region $\alpha<1$, providing a dynamical nonlocal (dNL) regime and demonstrating nonlocal effects, independent of both the local spin dimension \( d \) and the spatial dimension \( D \) for the Ising model. However, our analysis reveals that the maximal spreading of correlation (${C^d_2}^{\max}$) in fact depends on \( d \). Therefore, it can be argued that LR systems with $\alpha>1$ can serve as good quantum memories \cite{Roberts2020,Guo2024, zeng2024, Zeng2025} since the propagation of errors is bounded for a finite time, just like the cluster state generated via the NN interacting Hamiltonian \cite{Briegel2001}.

Note that we have also analyzed the dynamics of half-block entanglement entropy~\cite{dur2005,nielsenchuang, gong_gorshkov_prl_2017} and topological entanglement entropy~\cite{levin_wen_2006, kitaev_preskill_2006, Zeng2016, QImQM}. Their behaviors suggest monotonic dependence on the local Hilbert space dimension (see Appendices~\ref{app:entanglement_entropy} and ~\ref{app:TEE} for details).

 \begin{figure*}
    \centering
    \vspace{-3.0mm}
    \includegraphics[width=0.7\linewidth]{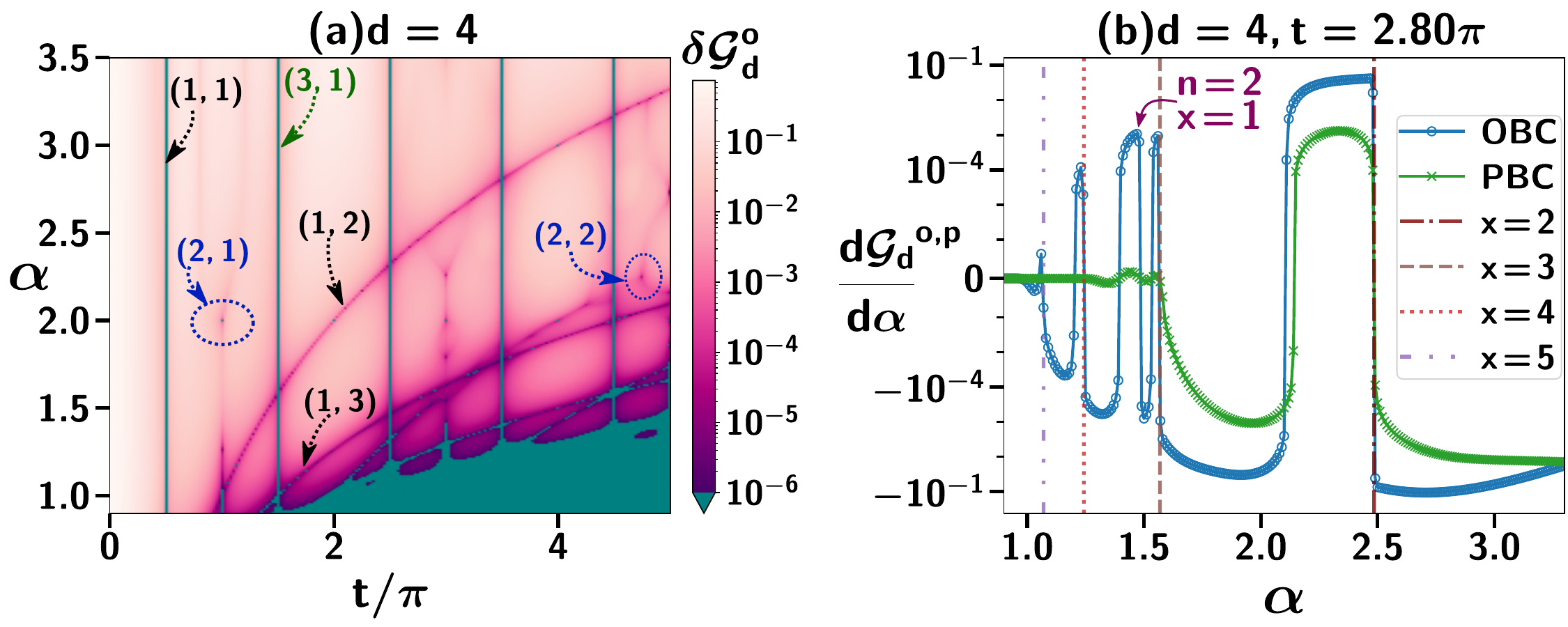}
    \vspace{-3.0mm}
    \caption{Non-analyticity of GGM, $\mathcal{G}^o$, for $d=4$ in the open chain. (a) $\delta\mathcal{G}^o_d=\mathcal{G}^{\max}_d-\mathcal{G}^o_d$ with respect to the fall-off rate $\alpha$ (ordinate) and time $t/\pi$ (abscissa). At time, $t_{n,x}=2\pi nx^\alpha/d, \delta\mathcal{G}^o_d=0$ ($<10^{-3}$ with $\Delta\alpha=\Delta t/\pi=10^{-2}$ spacings) shown in logarithmic scales. Various $(n,x)$-pairs are marked by arrows. (b) $\frac{\mathrm{d}\mathcal{G}_d}{\mathrm{d}\alpha}$ (ordinate) vs the fall-off rate $\alpha$ (abscissa) for PBC and OBC at $t=2.8\pi$. $\frac{\mathrm{d}\mathcal{G}_d^p}{\mathrm{d}\alpha}$ is continuous, while $\frac{\mathrm{d}\mathcal{G}_d^o}{\mathrm{d}\alpha}$ is discontinuous for $n=1$ and different $x$ with different vertical lines, and $n=2, x=2$ shown by an arrow. The ordinate is in symmetric-logarithmic scale ($10^{-5}$ linear threshold). Here $N=4000$, and all axes are dimensionless.}
    \label{fig:ggm_transition}
\end{figure*}

\section{Connecting LRB with genuine multipartite entanglement}
\label{sec:GGM}

An $N$-party pure state is genuinely multipartite entangled iff it is not separable across any bipartition. Let us now explore the potential of GME of the WGS to track the characteristics of LRB found in Sec.~\ref{sec:LRB}. Specifically, we use the generalized geometric measure (GGM) \cite{GGM_wei2003,ASD_GGM2010} to quantify the GME content of the dynamical pure WGS. For an $N$-party pure state $|\psi\rangle$ with vertices $V=\{1, 2,\dots, N\}$, the GGM, $\mathcal{G}$ is computed as $\mathcal{G}(|\psi\rangle) = 1-\max_{A}\{\lambda^2_{A:V-A}\}$. Here $\lambda_{A:V-A}$ represents the maximum Schmidt coefficient in the bipartition $A$ and $V-A$. Our explicit numerical analysis for WGS on a $1$D lattice demonstrates that the contribution in GGM arises exclusively from eigenvalues of single-site density matrices, i.e., \( A = \{i\} \) for \( i \in V \) (see Appendix~\ref{app:ggm} for details).  With PBC, eigenvalues of all density matrices are equivalent, while for OBC, in GGM,  the sites on the edge contribute, i.e., $A=\{1\}$ or $A=\{N\}$, suggesting GGM as an edge property of the WGS.

By denoting the GGM as $\mathcal{G}^{o(p)}_d(\alpha,t)$ for OBC (PBC), let us first discuss its important features in the case of NN interactions. In the qubit case, the NN Ising Hamiltonian ($\alpha\to\infty$) with OBC generates GGM for WGS which can be expressed as $\mathcal{G}^{o}_2(t)=\frac{1}{2}\big(1-\big|\cos\frac{t}{2}\big|\big)$ \cite{ghosh2023entanglement}. The expression clearly indicates that \(\mathcal{G}^{o}_2(t)\) attains its maximum and exhibits nonanalytic behavior with time at \( t_n = n\pi \) for \( n \in \mathbb{N} \), where \( n \) and \( 2 \) are coprime, i.e., \( n \) is odd. In contrast, for PBC, $\mathcal{G}_{2}^{p}(t) = \frac{1}{2}\big(1 - \cos^2\frac{t}{2}\big) \geq \mathcal{G}_{2}^o(t)$ (see Appendix~\ref{app:ggm}) which remains analytic at all times and achieves its maximum at $t_n$. Hence, for $d=2$, we have $\mathcal{G}_{2}^p(t_n) = \mathcal{G}_2^o(t_n) = \frac{1}{2} = \mathcal{G}_2^{\max}$. For higher $d$,  while the closed-form expressions of $\mathcal{G}^{p(o)}_{d}$ are hard to calculate, such dependence of GGM on the boundary condition with the NN spin-$s$ Ising Hamiltonian, is seen numerically (e.g., $d=3,4,10$) (shown in Appendix~\ref{app:ggm}). When local spin dimension increases, the relation $\mathcal{G}_d^{p}(t)\geq\mathcal{G}_d^{o}(t)$ holds, with equality occurring at $t_n=2n\pi/d$ ($n\in\mathbb{N}$). However, GGM reaches maximum value, $\mathcal{G}_d^{\max}=\mathcal{G}_{d}^p(t_n)=\mathcal{G}_d^o(t_n)=1-\frac{1}{d}$ only when $n$ and $d$ are coprimes, where WGS is the qudit cluster state and has the symmetry-protected topological order in $1$D \cite{clu_spt}. Note that the nonanalytic behavior of  $\mathcal{G}_d^o(t)$ at times other than $t_n$ is observed for $d>2$. For example, $\mathcal{G}_3^o(t)$ is non-analytic at $t_m=(m+\frac{1}{2})\pi$ ($m\in\mathbb{N}$) also, but $\mathcal{G}_3^o(t_m)\neq\mathcal{G}_3^{\max}$.

To find out the occurrences of $\mathcal{G}_d^{\max}$ in the presence of LR interactions, we analyze $\delta\mathcal{G}^o_d=\mathcal{G}^{\max}_d-\mathcal{G}^o_d$ [Fig.~\ref{fig:ggm_transition}(a) for $d=4$]. Along with the validity of $\mathcal{G}_d^{p}(\alpha, t)\geq\mathcal{G}_d^{o}(\alpha, t)$, our numerical assessment confirms that $\delta\mathcal{G}^o_d=0$ is satisfied if $t=t_{n,x}=2\pi nx^\alpha/d$ ($n\in \mathbb{N}$) when $n$ and $d$ are coprimes and $x\in\mathbb{N}$ for $1$D [$n=1,3$ for $d=4$ are illustrated in Fig.~\ref{fig:ggm_transition}(a), for different $\alpha$]. Note that, since $t_{n,x}/\pi$ is irrational at non-integer $\alpha$, we obatin $\delta\mathcal{G}^o_d<10^{-3}$ for $x>1$ with spacing of $10^{-2}$ between $\alpha$ and $t/\pi$ points. To confirm the position of $\mathcal{G}_d^{\max}$ in the $(t-\alpha)$ plane, we analyze the nonanalyticity of $\mathcal{G}^o_d$. To achieve the same, we fix $t=2.80\pi$ in ququart ($d=4$) scenario in Fig.~\ref{fig:ggm_transition}(b) and examine the behavior of  $\mathrm d\mathcal{G}^o_d/\mathrm{d}\alpha$ for $n=1$ (i.e., $n$ and $d$ are coprime) and $x=2,3,4$ and $5$, which reveals nonanalyticity at specific values of the fall-off rate satisfying $\alpha=\log_{x}(5.60/n)$ with $\delta\mathcal{G}^o_d= 0$. Even when \( d = 4 \) and \( n \) are not coprimes (e.g., \( n = 2 \)), the derivative remains nonanalytic at \( \alpha = \log_{x=2} 2.8 \), corresponding to \( \delta\mathcal{G}^o_d < 10^{-3} \), though such non-analyticity do not occur for all \( \alpha \). Finer numerical analysis of $t$ and $\alpha$ values reveals that  $\delta\mathcal{G}^o_d=0$ only at $t_{n,x}=2\pi nx^\alpha/d$ when $n$ and $d$ are coprimes.

Interestingly, the commutator $C^d_2(\alpha,\vec{r},t_{n,x})={C^{d}_2}^{\max}$, is maximum at the same time $t=t_{n,x}$ with $x=|\vec{r}|$, where $\delta\mathcal{G}^o_d=0$ with $n$ and $d$ as coprimes. Therefore, ${C^{d}_2}^{\max}$ necessarily indicates $\mathcal G_d^{\max}$ in our model. Thus, our investigation establishes a connection between the maximum possible geometric measure of GME content in WGS with OBC and its LR light cone, along with the subsequent maxima of the correlation function \( C^d_2(\alpha, \vec{r}, t) \). In Appendix~\ref{app:ggm_2d} we discuss the results of GGM of WGS defined on $2$D lattice geometry, where the same relation is obtained for GME, whereas $C^d_2(\alpha,\vec{r},t)$ is independent of lattice geometry.

\section{Conclusion}
\label{sec:conclusion}
The elegance of a one-way quantum computer lies in the notion of creating a highly entangled state in a physically realizable system on which a universal set of quantum gates can be implemented using local measurements.  Although the original proposal is for lattice containing spin-$1/2$ particles, it was extended to qudit cluster states, generated from higher-dimensional Ising-type interactions as the foundational framework for execution of  higher-dimensional measurement-based quantum computation (MBQC). While implementing MBQC in trapped ions or other physical systems, it is natural to consider an evolving Hamiltonian with variable-range interaction. 

This work demonstrates that the framework of weighted graph states, created through variable range interacting system, can be applied to reveal features of the evolving Hamiltonian based on the trends of the information-theoretic quantities, like mutual information (MI), maximum information spreading in Lieb-Robinson correlation, and genuine multipartite entanglement (GME). We established that the dependence of local dimension on the fall-off rate, dividing nonlocal and local regimes, predicted by MI can be confirmed by LRBs and GME. Our results indicate that the notion of cluster-like states can be applied to probe the nonequilibrium physics in many-body systems.

\section{Acknowledgments}

We acknowledge the support from Interdisciplinary Cyber Physical Systems (ICPS) program of the Department of Science and Technology (DST), India, Grant No. DST/ICPS/QuST/Theme- 1/2019/23. We  acknowledge the use of \href{https://github.com/titaschanda/QIClib}{QIClib} -- a modern C++ library for general purpose quantum information processing and quantum computing~\cite{qiclib} and cluster computing facility at Harish-Chandra Research Institute. KDA and PH acknowledges ``INFOSYS scholarship for senior students".

% \onecolumngrid
%\section{Appendix}
\appendix
% The appendices contains a detailed discussions on --- (1) the calculation of reduced density matrices of the weighted graph state in one-dimensional ($1$D) lattice produced through the dynamics governed by variable range interacting Hamiltonian; (2) the scaling of its entanglement entropy; (3) \& (4) behavior of topological entanglement entropy and mutual information with time; (5) computation of Lieb-Robinson bound in WGS; (6) analysis of genuine multipartite entanglement (GME) in this state; and finally, (7) the study of GME in two dimensional lattices.

\begin{figure*}
    \centering
    \vspace{-3.0mm}
    \includegraphics[width=0.8\linewidth]{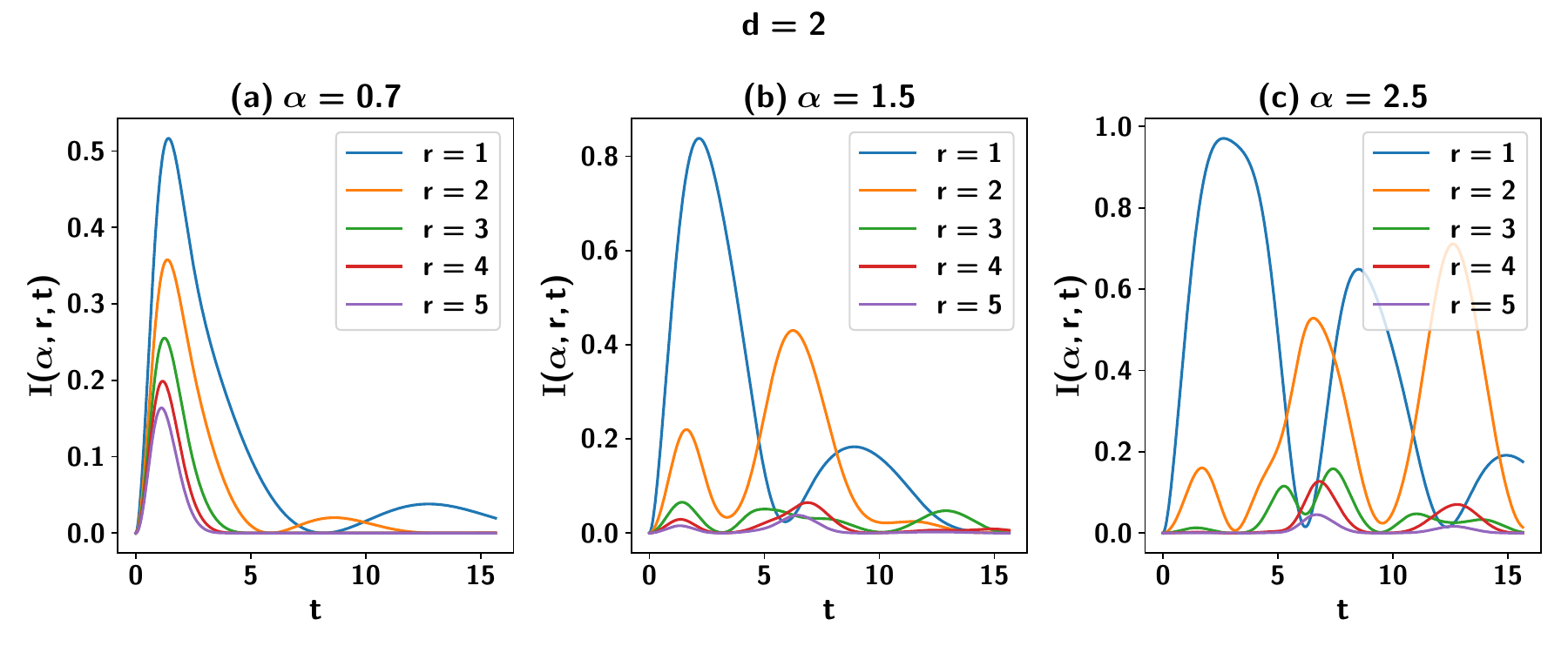}
    \vspace{-7.0mm}
    \caption{Mutual information, $I(\alpha,r,t)$ (ordinate) against time $t$ (abscissa) for the various fall-off rates $\alpha=0.7,1.5,2.5$ (a)--(c) for a spin-$1/2$ Ising Hamiltonian on a $1$D lattice used for evolution.  Here  $N=10^3$. Different lines represent different values of separation $r$ between two qubits in the lattice. Both the axes are dimensionless.} 
    \label{fig:i_rt_sig_d2}
\end{figure*}
\section{Reduced density matrices of weighted graph states}
\label{app:reduce_density_matrix}

We use the projected entangled pair state (PEPS) formalism \cite{cirac2004,dur2005,Hartmann2007} to efficiently compute the reduced density matrices of the WGS on vertices $V=\{1, 2, \dots,N\}$, generated by the long-range spin-$s$ Ising Hamiltonian $H_d(\alpha)$, as given in Eq. (1) of the main text, where each summand is commuting. Therefore, the time evolution is represented by product of commuting unitaries, i.e, $U = \prod_{x<y\in V}U_{xy}$, with $U_{xy}=\sum_{a_x, a_y=0}^{d-1}\exp \!\left(\! -ia_x a_y \frac{t}{|\vec{r}_{xy}|^{\alpha}}\!\right)\! \ket{a_x a_y}\bra{a_x a_y}$, where $|\vec{r}_{pq}|$ denotes the distance between the sites $p$ and $q$, with a fall-off rate $\alpha$. Let us divide the whole $N$-site lattice into two subsystems, $A$ and $\bar A$ where $\bar A$ is the complement of $A$, i.e.,, $A\cup \bar{A}$ constitutes the whole system. Note that the unitaries acting only on the qudits in $\bar{A}$ does not effect the subsystem $\rho_{A}$. The commuting unitaries acting on each site from $A$ and $\bar A$ lead to a state
\begin{eqnarray}
\ket{\psi'} = \prod_{p\in A,q\in \bar{A}}U_{pq}\ket{+}^{\otimes N},
\end{eqnarray}
from which we get a reduced state as $\rho'_A = \Tr_{\bar{A}}\ket{\psi'}\bra{\psi'}$. We can write the reduced state of subsystem $A$, denoted by $\rho_A$ as
\begin{eqnarray}
    \rho_A = \prod_{p_1,p_2\in A}U_{p_1p_2}\rho'_A U_{p_1p_2}^\dagger.
\end{eqnarray} 
Note that the eigenvalues of $\rho'_A$ are the same as those of $\rho_A$.

In the PEPS picture, the qudit on vertex $x\in V$ of an $N$-qudit state is replaced by $N-1$ virtual qudits $x_y$ ($y\in V-\{x\}$). Here, for $p\in A, q\in\bar{A}$, the virtual qudit $q_p$ of $q$ is connected with the virtual qudit $p_q$ of $p$ by the unitary $U_{pq}$. To return back to the physical qudit, projectors $P_x$ of the form
\begin{eqnarray}
    P_x=\sum_{k=0}^{d-1}\ket{k}_x\bra{k_1,k_2,\ldots, k_{x-1},k_{x+1},\ldots, k_{N}},\label{eq:proj}
\end{eqnarray}
are applied first on each vertex $x\in\bar{A}$. We then obtain the state where all the physical qudits of $\bar{A}$ are entangled with the virtual qudits of $A$. For each of the virtual entangled states, the reduced density operator $\rho'_A(q)$ of $A$ corresponding to each $q\in \bar{A}$ can be obtained by tracing out the $q^{th}$ vertex. Then the reduced density matrix can be written as
\begin{eqnarray}
    \rho'_A(q) &=& \frac{1}{d}\sum_{\nu=0}^{d-1}\ket{\phi_q^\nu}\bra{\phi_q^\nu}\ ;\nonumber\\
    \ket{\phi_q^\nu} &=& \bigotimes_{p\in A}\sum_{\mu_p=0}^{d-1}\frac{e^{i\mu_p\nu t/|\vec{r}_{pq}|^\alpha}}{\sqrt{d}}\ket{\mu_p}.
\end{eqnarray}
Using Eq.~\eqref{eq:proj} for the qudits \( p \in A \) results in the Hadamard product of all \( \rho'_A(q) \) for \( \forall q \in \bar{A} \), yielding \( \rho'_A \). For example, the single-site reduced density matrix for the first site takes the form
\begin{equation}
    [\rho_{\{1\}}]_{\mu\nu}=\frac{1}{d}\prod_{q=2}^N\bigg(\frac{1}{d}\sum_{k=0}^{d-1}\exp[i(\mu-\nu)k \:\frac{t}{|\vec{r}_{1q}|^\alpha}]\bigg),
    \label{eq:rho1}
\end{equation}
with $\mu,\nu\in\{0, 1,\dots,d-1\}$. This expression is useful in the calculation of the multipartite entanglement of the WGS, as shown in further sections. In a similar fashion, it is possible to obtain all the reduced density matrices of the dynamical state after making an arbitrary partition in the system. These steps are also explained with examples in \cite{ghosh2023entanglement} for qubits ($d=2$). Note, however that to calculate any quantum information theoretic quantities, we shall need to diagonalize these matrices, which require exponentially (with the size of the subsystem) large computational resources. 

\begin{figure*}
    \centering
    \vspace{-3.0mm}
    \includegraphics[width=0.8\linewidth]{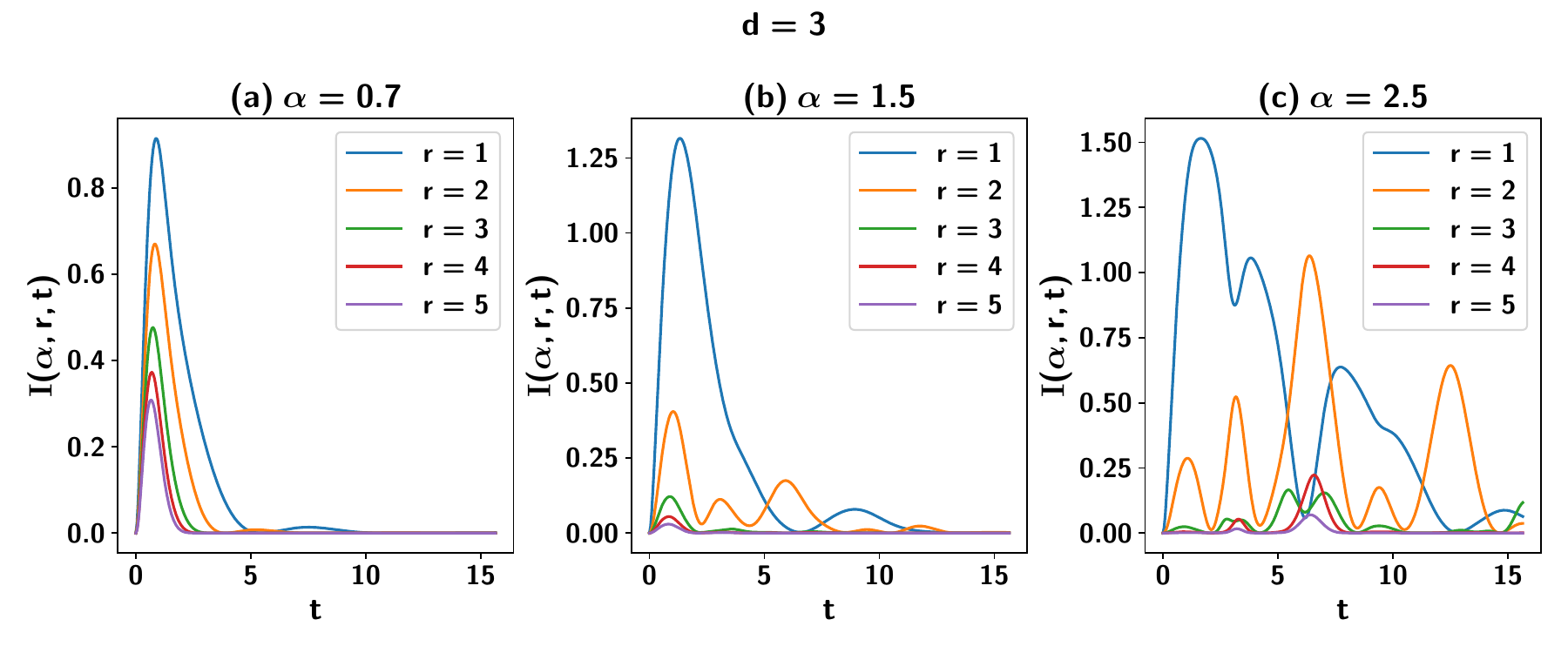}
    \vspace{-7.0mm}
    \caption{All specifications remain the same as in Fig.~\ref{fig:i_rt_sig_d2}, except that qubits are replaced with qutrits (\( d = 3 \)).} 
    \label{fig:i_rt_sig_d3}
\end{figure*}

\section{Oscillation in mutual information for multiqudit WGS}
\label{app:mutual_info}

We see the trends of mutual information (MI) $I(A:B)$ between sites $A=\{1\}$ and $B=\{r+1\}$, separated by a distance $r$ in the main text, and we characterize the logarithmic connection of fall-off rate $\alpha$ with the local spin dimension $d$ on the spread of mutual information in the WGS in $1$D with an open boundary condition. Here we explicitly show the dynamical profile of mutual information in WGS for various fall-off rates $\alpha$ and local spin dimension $d$.

In both Figs. \ref{fig:i_rt_sig_d2} and \ref{fig:i_rt_sig_d3}, for a small fall-off rate $\alpha$, MI is significant at small distances ($r=1$ to $5$) only at small time, suggesting the spread of $I$ to larger distances after certain times, while for large fall-off rates $\alpha$, MI remains high at small distances even for larger times. The profile of $I$ for $r=4, 5$ also shows small values at a large fall-off rate $\alpha$, thereby encouraging us to study the scaling of mutual information $I$ with distance. Moreover, as depicted in the main text, the time-averaged MI changes its behavior from quadratic to linear with distance between two subsystems approximately at $\alpha\sim \log_2d$, highlighting the dependency of local Hilbert space at the transition point. Note that, on $1$D lattice with PBC, we find that while a similar profile of $\log \langle I \rangle_{t_0}$ exists with $r$, such a clear logarithmic dependence on fall-off rate $\alpha$ is absent. We can predict this by analyzing LRBs and GME content of the dynamical state.

\begin{figure*}
    \centering
    \includegraphics[width=\linewidth]{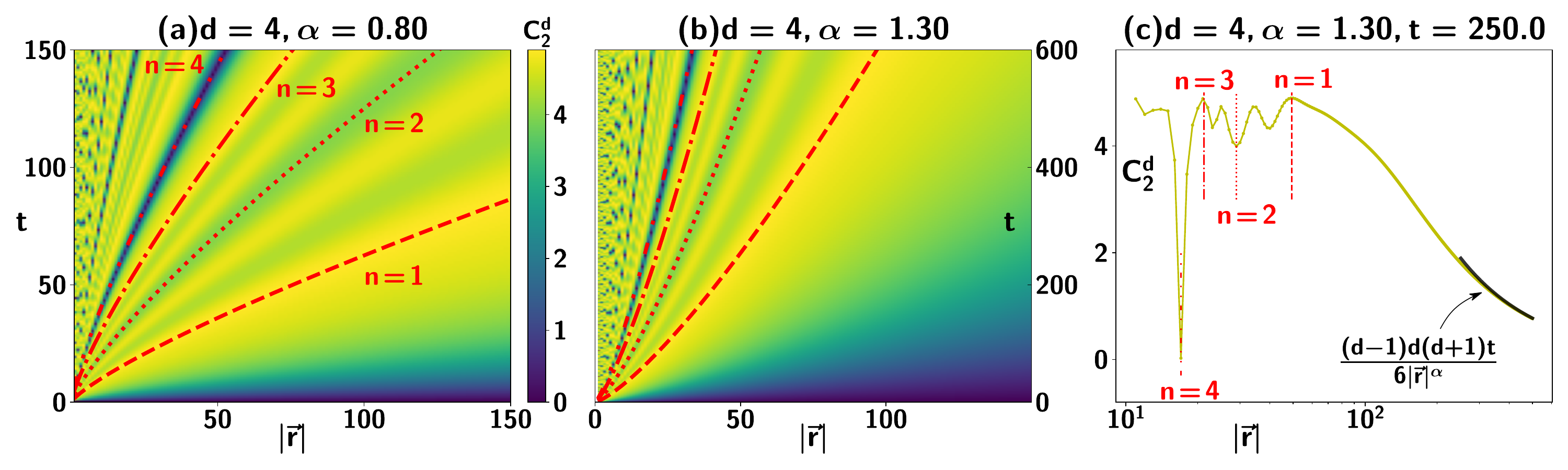}
    \caption{Information spread for $d=4$ in the qudit LR Ising model $H_d(\alpha)$, quantified by the commutator $C^d_2$ (colorbar) for fall-off rates (a) $\alpha=0.8$ (b) $\alpha=1.3$ in the time $t$ (ordinate)--distance $|\vec{r}|$ (abscissa) plane. The red lines represent the $t_{n,|\vec{r}|}=2\pi n|\vec{r}|^\alpha/d$ for $n=1$ (dashed), $n=2$ (dotted), $n=3$ (dash-dotted) and $n=4$ (dashdot-dotted) (c) $C^d_2$ (ordinate) for $d=4$ at $\alpha=1.3$ at fixed time $t=250$ against distance $|\vec{r}|$ (abscissa). Maximum $C^{d \max}_2$ was obtained at $t_{n,|\vec{r}|}$ only for $n=1,3$ and minimum for $n=d=4$. The tail, i.e., $|\vec {r}|\gg (td/2\pi)^{\frac{1}{\alpha}}$ gives $C^d_2(\alpha,\vec{r},t)\sim (d-1)d(d+1)t/|\vec{r}|^\alpha$ (black solid line).  All the axes are dimensionless.} 
    \label{fig:lrb_d4_c2}
\end{figure*}

\section{Calculation of Lieb-Robinson bounds in WGS}
\label{app:lrb}
Lieb-Robinson bounds are used to understand the speed of the propagation of information in non-relativistic quantum systems. Let the initial state, \( \ket{\psi_{\text{in}}} \), evolve under the Hamiltonian \( H \). Two local operators, \( A_{\vec{x}}=\sum_{a_x, b_x=0}^{d-1} A_{b_x}^{a_x}\ket{b_x}\bra{a_x} \) and \( B_{\vec{y}}=\sum_{a_{y}, b_{y}=0}^{d-1} B_{b_y}^{a_y}\ket{b_y}\bra{a_y} \), are defined on the lattice sites \( x \) and \( y \), with position coordinates \( \vec{x} \) and \( \vec{y} \), respectively. Due to a small perturbation $\epsilon$ on site $y$, the effect on site $x$ is captured by the change in dynamics of the time-evolved operator $A_{\vec{x}}(t)=e^{-iHt}A_{\vec{x}}e^{iHt}$ in the Heisenberg picture. Mathematically, it can be written as 
\begin{eqnarray}
    \hspace{-20.0mm} \bra{\psi_{\text{in}}}&\hspace{-6.0mm} e^{i\epsilon B_{\vec{y}}} A_{\vec{x}}(t) e^{-i\epsilon B_{\vec{y}}} \ket{\psi_{\text{in}}} - \bra{\psi_{\text{in}}} A_{\vec{x}}(t) \ket{\psi_{\text{in}}} \nonumber\\ &= -\epsilon \bra{\psi_{\text{in}}} A_{\vec{x}}(t) B_{\vec{y}} -B_{\vec{y}} A_{\vec{x}}(t) \ket{\psi_{\text{in}}} +O(\epsilon^2).
\end{eqnarray}
Therefore, for small $\epsilon$, the operator norm [Schatten-$p$ ($p\geq1$) norm] of the commutator $[A_{\vec{x}}(t), B_{\vec{y}}]$ is of interest which gives the bound on information propagation in quantum systems for any initial state. Specifically, $||[A_{\vec{x}}(t), B_{\vec{y}}]||_p$ measures the growth of $A_{\vec{x}}(t)$ on the support of operator $B_{\vec{y}}$ which is zero at $t=0$. It has been studied extensively in local and gapped systems, as well as long-range systems to quantify and bound the speed of propagation of information \cite{Eisert2013,Gong2014,Foss-Feig2015,Tran2019,chen2019,Else2020,Yin2020,kuwahara2020,tran2020,Tran2021_LRB,Chen2021,Gong2023,lemm2024EnLRB}.

In this work, we consider the $N$-qudit WGS, with vertex set $V=\{1,2,\dots, N\}$ and $x,y\in V$, where the evolution is governed by the spin-$s$ long-range Ising Hamiltonian $H_d(\alpha)$. The effect of $H_d(\alpha)$ on the computational basis is described as $H_d(\alpha)\ket{\vec{a}} = \mathcal{E}_{\vec{a}}(\alpha)\ket{\vec{a}}\bra{\vec{a}}$, as shown in the main text [Eq.$(1)$], with $\ket{\vec{a}}\equiv\ket{a_1, a_2,\dots, a_N}$ forming the computational basis ($a_k\in\{0, 1, \dots, d-1\}$). The energy $\mathcal{E}_{\vec{a}}(\alpha)=\sum_{x<y\in V}a_x a_y/|\vec{r}_{xy}|^\alpha$ is the eigenvalue of eigenstate $\ket{\vec{a}}$ with $\vec{r}_{xy}=\vec{y}-\vec{x}$ on any lattice.

For notational simplicity, we denote \( \ket{\vec{a}[b_x, b_y]} \) (where \( b_x, b_y \in \{0,1,\dots, d-1\} \)) as the state in which the sites \( x \) and \( y \) take the values \( b_x \) and \( b_y \), respectively, while all other sites \( k \) (\( k \in V, k \neq x, y \)) assume the default values \( a_k \). Note that, according to this notation, $\ket{\vec{a}[a_x, a_y]}\equiv\ket{\vec{a}}$ and $\ket{\vec{a}[b_x, a_y]}\equiv\ket{\vec{a}[b_x]}$. Therefore, in the $d^N$ computational basis,
\begin{eqnarray}
    A_{\vec{x}}(\alpha,t) &=& \sum\limits_{\vec{a}}\sum_{b_x=0}^{d-1} A_{b_x}^{a_x}\exp\bigg(it\Delta\mathcal{E}^{b_x}_{a_x}(\alpha)\bigg)\ket{\vec{a}[b_x]}\bra{\vec{a}}, \nonumber\\
    B_{\vec{y}} &=& \sum\limits_{\vec{a}}\sum_{b_y=0}^{d-1}B_{b_y}^{a_y}\ket{\vec{a}[b_y]}\bra{\vec{a}}, %\text{ where }\\
\end{eqnarray}
where
\begin{eqnarray}
\Delta\mathcal{E}^{b_x}_{a_x}(\alpha)  &=& \mathcal{E}_{\vec{a}[b_x]}(\alpha) - \mathcal{E}_{\vec{a}}(\alpha) \nonumber \\
&=& (b_x-a_x)\sum_{\mathclap{j\in V-\{x\}}}a_j/|\vec{r}_{xj}|^\alpha.
\end{eqnarray}
Now we explicitly calculate the above-mentioned commutator as
\begin{widetext}
\begin{eqnarray}
   \nonumber &&[A_{\vec{x}}(\alpha,t), B_{\vec{y}}]=\sum\limits_{\vec{a}}\sum_{b_x, b_y=0}^{d-1}A_{b_x}^{a_x}B_{b_y}^{a_y} \Bigg( \exp(it\Delta\mathcal{E}^{b_x,b_y}_{a_x, b_y}(\alpha))-\exp(it\Delta\mathcal{E}^{b_x}_{a_x}(\alpha))\Bigg) \ket{\vec{a}[b_x,b_y]}\bra{\vec{a}}\nonumber\\
   &&=\sum\limits_{\vec{a}}\sum_{b_x, b_y=0}^{d-1} \!\!\!A_{b_x}^{a_x}B_{b_y}^{a_y} \exp(\!it(b_x-a_x)\sum_{\mathclap{j\in V-\{x,y\}}}a_j/|\vec{r}_{xj}|^\alpha\!)\!\Bigg(\!\!\exp(\!ib_y(b_x-a_x)\frac{t}{|\vec{r}_{xy}|^\alpha}\!) - \exp(\!ia_y(b_x-a_x)\frac{t}{| \vec{r}_{xy}|^\alpha}\!)\!\!\Bigg)\!\ket{\vec{a}[b_x,b_y]}\bra{\vec{a}} \nonumber
\end{eqnarray}
\begin{eqnarray}
    &&=\tilde{U} \Biggl\{ \sum\limits_{\vec{a}}\sum_{b_x, b_y=0}^{d-1}A_{b_x}^{a_x}B_{b_y}^{a_y}\Bigg(\exp(ib_y(b_x-a_x)\frac{t}{| \vec{r}_{xy}|^\alpha}) - \exp(ia_y(b_x-a_x)\frac{t}{| \vec{r}_{xy}|^\alpha})\Bigg)\ket{\vec{a}[b_x,b_y]}\bra{\vec{a}} \Biggr\} \tilde{U}^{\dagger},
\end{eqnarray}
\end{widetext}
where $\tilde{U} = \sum_{\vec{a}} \exp(it\:a_x) \sum_{j\in V-\{x,y\}} a_j/|\vec{r}_{xj}|^\alpha) \ket{\vec{a}} \bra{\vec{a}}$ is a diagonal unitary. The term inside $\bigl\{\cdot\bigl\}$ is a block-diagonal matrix with repeated $d^2\times d^2$ blocks. Therefore, $\tilde{U}^{\dagger}[A_{\vec{x}}(\alpha,t), B_{\vec{y}}]\tilde{U}$, despite being of dimension $d^N\times d^N$, have only $d^2$ unique singular values (in singular value decomposition). Therefore, without loss of generality, site $x$ is at origin $\vec{x}=\vec{0}$ and site $y$ is at $\vec{y}=\vec{r}$. We define the $d^2\times d^2$-dimensional operator $\bar{C}(d, \alpha, \vec{r},  t)$ having elements as 
\begin{equation}
    \bar{C}^{a_x,a_y}_{b_x,b_y} \!=\! \left(\!e^{ib_y(b_x\!-a_x)\!\frac{\!\!t}{|\! \vec{r}|^{\!\alpha}}} - e^{ia_y(b_x\!-a_x)\!\frac{\!\!t}{|\! \vec{r}|^{\!\alpha}}}\!\right)\! A^{a_x}_{b_x} B^{a_y}_{b_y}.
\end{equation}
\begin{figure*}
    \centering
    \vspace{-3.0mm}
    \includegraphics[width=0.7\linewidth]{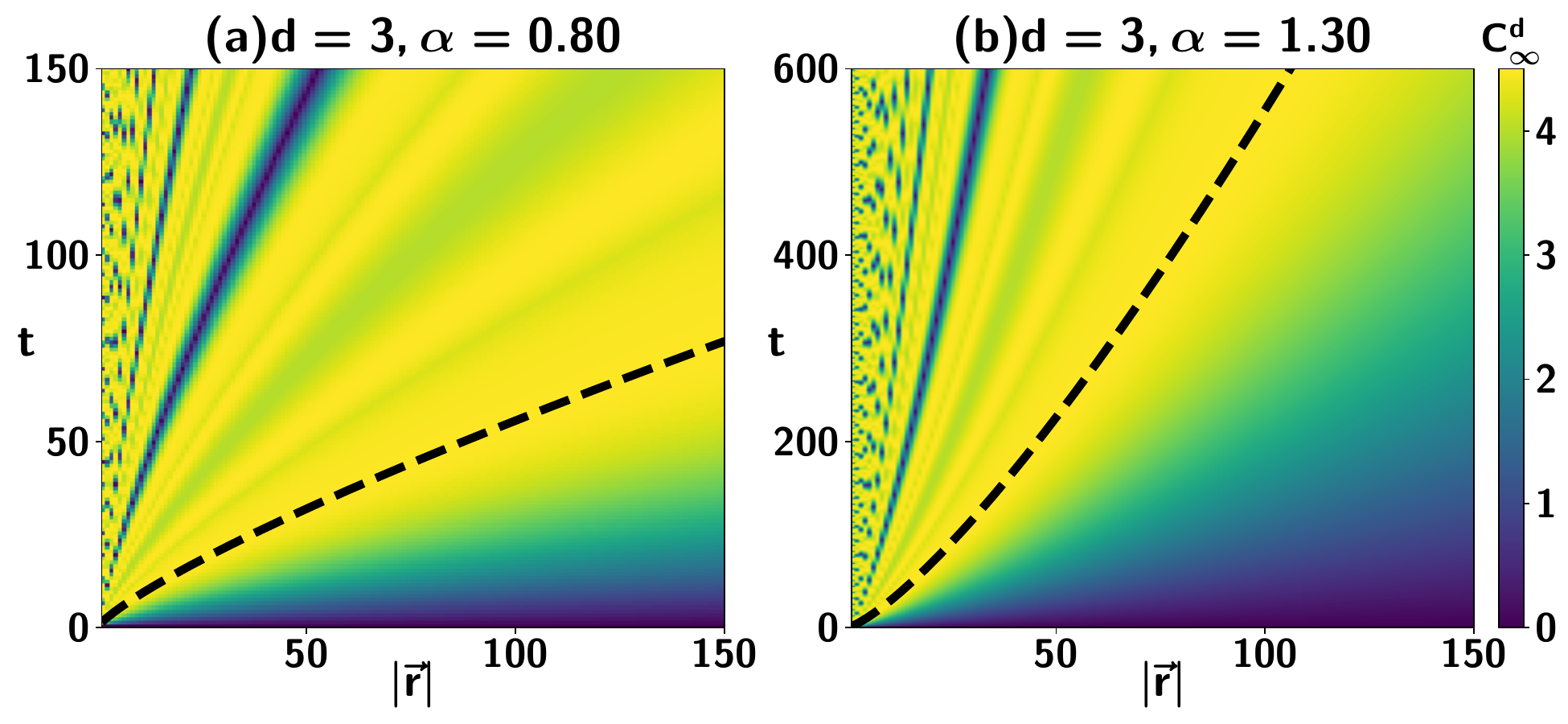}
    \vspace{-5.0mm}
    \caption{Information spread for $d=3$ in the qudit LR Ising model $H_d(\alpha)$, quantified by the commutator $C^d_\infty$ (colorbar) for fall-off rates (a) $\alpha=0.8$ (b) $\alpha=1.3$ in the time $t$ (ordinate)--distance $|\vec{r}|$ (abscissa) plane. The dashed (black) line represents the $t_{|\vec{r}|}=4\pi|\vec{r}|^\alpha/d^2$ line . All the axes are dimensionless.} 
    \label{fig:lrb_dinf}
\end{figure*}
All the singular values of \( [A_{\vec{x}}(\alpha,t), B_{\vec{y}}] \) are identical to those of \( \bar{C}(d, \alpha, \vec{r}, t) \), with each singular value having a \( d^{N-2} \)-fold degeneracy. We are now ready to calculate $\lVert [A_{\vec{x}}(\alpha,t), B_{\vec{y}}] \rVert_p = C^d_p(\alpha, \vec{r}, t)$ for any valid $p$. As given for $p=2$ in the main text, $C^d_p(\alpha, \vec{r}, t)$ is maximum for \( A = B = d\ket{+}_d\bra{+} \), which is maximum, i.e., $C_2^d(\alpha,\vec{r},t)=C_2^{d\max}$ at $t_{n,|\vec{r}|}=2\pi n|\vec{r}|^\alpha/d$, only when $n$ and $d$ are coprimes, as shown in Fig. \ref{fig:lrb_d4_c2} for $d=4$, and this is also the case for higher $d$ (checked upto $d=10$). Figure \ref{fig:lrb_d4_c2} also shows $C_2^d=0$, i.e., resets at $t_{n,|\vec{r}|}$ when $n=4$, and subsequently, when $n=md$, for given $m\in\mathbb{N}$, giving the condition of $t_{m,|\vec{r}|}=2\pi m|\vec{r}|^\alpha$ for $C_2^d(\alpha, \vec{r}, t)=0$. Finally, far away from the light cone, i.e., at a given time $t$, and distance $|\vec{r}|\gg(td/2\pi)^{\frac{1}{\alpha}}$, $C_2^d(\alpha, \vec{r}, t)=\frac{(d-1)d(d+1)t}{6|\vec{r}|^\alpha}$, giving the polynomial fall-off. Therefore, the LRB for the long-range systems is obtained as shown previously in Ref. \cite{lemm2024EnLRB}.

Let us consider the same operators, \( A = B = d\ket{+}_d\bra{+} \), which maximize the \( p = 2 \) case, for the \( p \to \infty \) norm as well. The norm \( C_{\infty}^d(\alpha, \vec{r}, t) \) corresponds to the maximum singular value of \( \bar{C}(d, \alpha, \vec{r}, t) \) and attains its upper bound of \( d^2/2 \) at \( t_{|\vec{r}|} = 4\pi |\vec{r}|^\alpha / d^2 \), as illustrated in Fig. \ref{fig:lrb_dinf}. While the results obtained are similar with $\alpha = 2\log_{|\vec{r}|}d+\log_{|\vec{r}|}(t_{|\vec{r}|}/4\pi)$ and $C^d_{\infty}(\alpha, \vec{r^\prime}, t_{|\vec{r}|})\sim (d-1)d^2(d+1)t_{|\vec{r}|}/|\vec{r}|$, they cannot connect the observations from GGM like $C^d_p$ for $p=2$. These results show that the maximization over operators is probably needed for $p\to\infty$. 
The connection of LRB with $p=2$ to GGM of the WGS formalism may also be due to the fact that the Schatten $2$-norm is self-dual.

\begin{figure*}
    \centering
    \includegraphics[width=\linewidth]{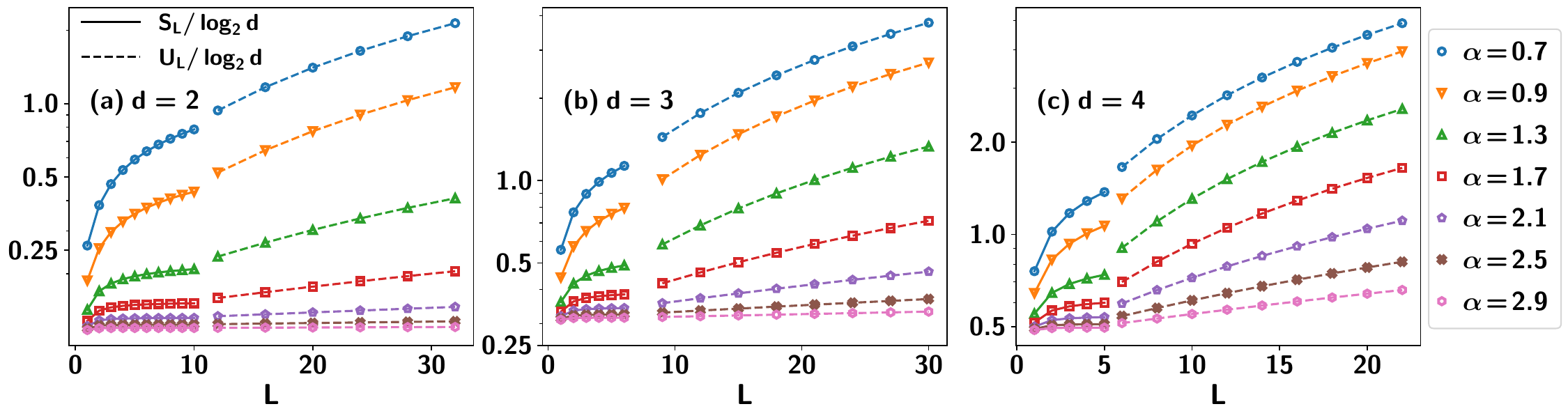}
    \caption{Normalized block entanglement entropy, $S_L/\log_2d$ and its upper bound, $U_L/\log_2d$ (ordinate in logarithmic scale) with block length $L$ (abscissa). Here system size $N=10^3$ and  $t=0.5$. The exact values of $S_L/\log_2d$ upto $L=10,6,5$ for local spin dimension $d=2, 3$, and $4$ are given by solid lines, whereas dashed lines represent the upper bound, $U_L/\log_2d$. The values of $\alpha$  up to which the area law can be found increase with increasing $d$. Note that  $U_L$ is calculated by taking $L/k=5, 3$, and $2$ for $d=2, 3$, and $4$ respectively. All the axes are dimensionless.}
    \label{fig:ent_detection}
\end{figure*}
\section{Scaling of entanglement entropy  for WGS}
\label{app:entanglement_entropy}

Given an arbitrary $N$-qudit system, we divide the system into two parts $A=\{1,2,\ldots,L\leq N\}$ and its complement $\bar{A}=V-A$. For a pure state, the block entanglement entropy between these two subsystems is defined as $S_L\equiv S(\rho_A) = S(\rho_{\bar{A}})$, where $\rho_A$ and $\rho_{\bar A}$ denote the reduced density matrices of $A$ and $\bar A$ respectively. Here, $S(\rho)=-\Tr(\rho \log_2 \rho)$ is the von Neumann entropy (vN) of $\rho$. In case of one-dimensional spin models, under \textit{volume law}, $S_L$ scales linearly with $|A|=L$, i.e., $S_L\sim L$ whereas $S_L\sim \log L$ corresponds to the \textit{sub-volume law}. $S_L$ remains constant when \textit{area law} is followed. The dynamics of entanglement entropy sheds light on the nonlocal properties of the quantum systems~\cite{dur2005, gong_gorshkov_prl_2017}.

Let us analyze the characteristics of normalized block entanglement entropy ($S_L/\log_2d$) in the case of WGS in $1$D (see Fig.~\ref{fig:ent_detection}). Specifically, we calculate the exact values of $S_L$ up to $L=10,6,$ and $5$ for $d=2,3$ and $4$ respectively, above which we calculate the upper bound of block entanglement entropy given by $S_L\leq \sum_{j=1}^{k-1}S(\rho_{A_j\cup A_{j+1}})-\sum_{j=2}^{k-1}S(\rho_{A_j})\equiv U_L$ using sub-additivity property of vN (see Ref. \cite{dur2005} in this respect). Note that, to calculate the bound, we divide the subsystem $A$ of length $L$ into $k$ equal parts $A_j$ of length $L/k$, such that $S(\rho_{A_j\cup A_{j+1}})$ is numerically computable. 

In the spin-$1/2$ system, we get the signature of area law for $\alpha>1.3$ since up to $L=10$, $S_L$ seems to saturate with $L$ at a small time $t=0.5$. Also, for $L>10$ the upper bound of $S_L$ increases linearly with $L$ for different values of  $\alpha<1.3$, hence a probable volume law. As $d$ increases, this transition of area law to volume law appears to be dependent on the local spin-dimension as shown in Fig.~\ref{fig:ent_detection}. The slope of all the curves of both $S_L$ and $U_L$ seems to be increasing with the increase of $d$. Also, we observe that the gap between $S_L$ and its upper bound $U_L$ corresponding to $\alpha=1.3$ and $0.9$ decreases with increasing the value of local spin-dimension. These observations suggest that the $d$ dependence of the transition from area law to volume law on $\alpha$ exists for the variable-range spin-$s$ Ising model, although such a dependence cannot be precisely obtained through the dynamics of entanglement entropy.

\begin{figure*}
    \centering
    \vspace{-3.0mm}
    \includegraphics[width=0.7\linewidth]{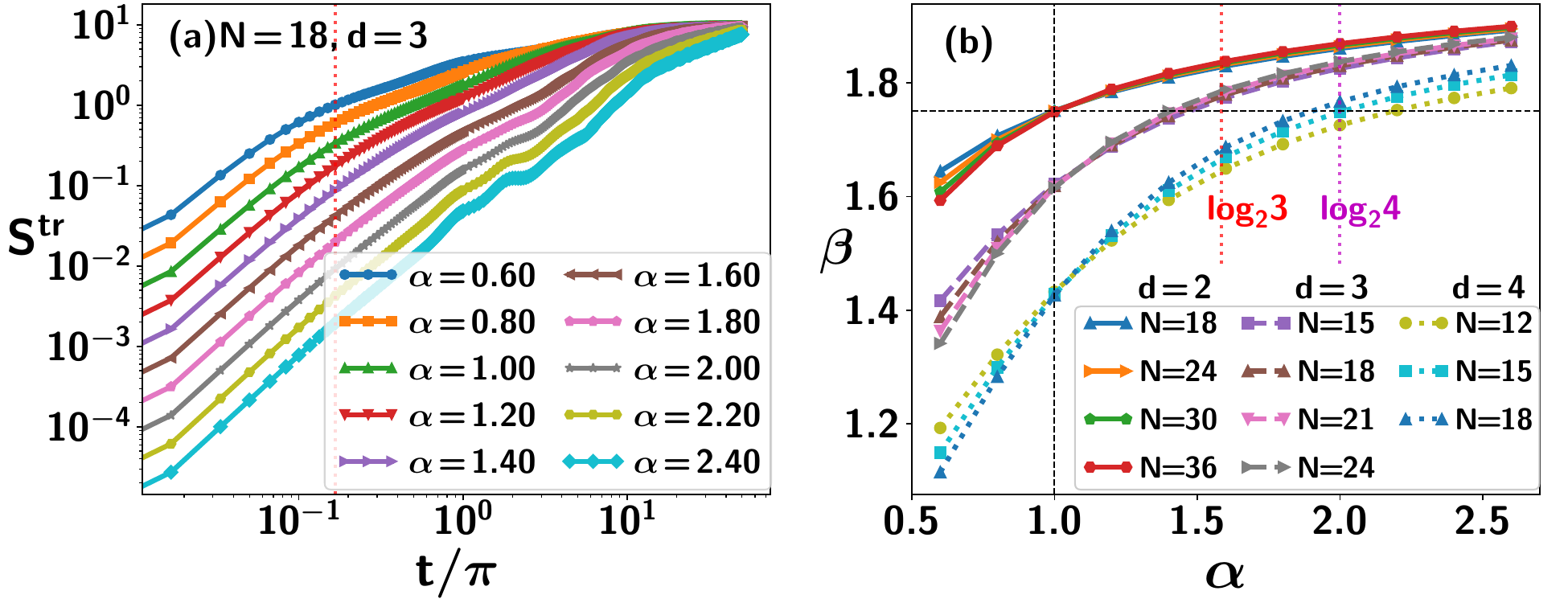}
    \vspace{-3.0mm}
    \caption{Dependence of TEE ($S^{\text{tr}}$ of WGS on $1$D open chain) growth with time on local spin dimension $d$. (a) TEE $S^{\text{tr}}$ (ordinate) against time $t$ (abscissa) for different fall-off rates $\alpha$ (with axes in logarithmic scales). (b) Exponent $\beta$ (ordinate) for fitting $S^{\text{tr}}=c t^\beta$ from $t>0$ to $t=\pi/6$ [dotted line in (a)], against $\alpha$ (abscissa), for different system size $N$ and local spin dimension $d$. While $\beta$ at different $N$ crosses at $\alpha=1$, the same value of $\beta$ at $\alpha=1, d=2$ is reached at $\alpha\sim\log_{2}d$ for $d=3$ and $4$. All the axes are dimensionless.} 
    \label{fig:st_topo}
 \end{figure*}

\section{Effect of long-range and local spin dimension on the topological properties: Topological entanglement entropy} 
\label{app:TEE}

Topological order is a nonlocal phenomenon which shows robustness to local perturbations and is quantified by topological entanglement entropy (TEE). Generically studied in the ground states of the local Hamiltonians, we use it to study the nonlocal properties in the dynamical WGS, prepared from the Ising Hamiltonian with LR interactions, $H^d(\alpha)$ in $1$D with open ends. While the system size scaling of entropy of the WGS at small times indicates volume-law behavior with a smaller fall-off rate, $\alpha$ in higher $d$ as shown in Appendix~\ref{app:entanglement_entropy}, its characterization is hindered by the small system sizes.

We use tripartite mutual information $S^{\text{tr}}$, which is computed by dividing the open chain into three equal parts, with $A$ and $C$ at the edges and $B$ in the center. TEE is defined as $S^{\text{tr}} = I(A:C|B) = S(\rho_{AB}) + S(\rho_{BC}) - S(\rho_{B}) + S(\rho_{ABC})$ where $S(\sigma)=-\Tr(\sigma\log_2\sigma)$ is the von Neumann entropy. Since the generated WGS is a pure state, $S(\rho_{ABC})=0$, $S(\rho_{AB})=S(\rho_{C})$ and $S(\rho_{BC})=S(\rho_{A})$. $S^{\text{tr}}$ in this partition removes the area-dependent terms in the entropy scaling, therefore quantifying volume law and sub-area law effects and it is known to distinguish both symmetry-breaking quantum phase transitions and topological phase transitions in the ground state of local gapped systems \cite{levin_wen_2006, kitaev_preskill_2006, Zeng2016, QImQM}.

The cluster state is a symmetric ground state of the local cluster Hamiltonian, which has global symmetry and the ground state degeneracy in the open $1$D lattice ($\mathbb{Z}_2\times\mathbb{Z}_2$ symmetry and four-fold degeneracy in qubits \cite{Zeng2016, QImQM}). It shows nonzero $S^{\text{tr}}$ only in the presence of symmetry-breaking term, while the symmetric ground state, which is also the case in the dynamical framework, has $S^{\text{tr}}=0$. Therefore, $S^{\text{tr}}$ of WGS in the dynamics of the LR Ising model shows the effects of the LR interactions. 

$S^{\text{tr}}$ in WGS increases with time $t$, as shown in Fig.~\ref{fig:st_topo}, and initially shows a power law growth as $S^{\text{tr}}=ct^\beta$, with the parameters $c$ and $\beta$ dependent on fall-off rate $\alpha$. In the transient regime, $S^{\text{tr}}$ starts to oscillate before converging to a value (which is found to be system size independent). Fitting the numerically obtained $S^{\text{tr}}$ for various $\alpha$, we see that the time exponent $\beta\equiv \beta_d(\alpha)$ shows two interesting properties. First, for any local spin dimension $d$, $\beta$ for various system sizes $N$ crosses at $\alpha=1$. Second, $\beta_{2}(1) \sim \beta_{3}(\log_{2}3)\sim \beta_{4}(\log_{2}4)$, which suggests a logarithmic dependence of the topological properties of WGS on local spin dimension. This analysis becomes difficult for higher $d$, as only $N\leq15$ system sizes can be accessed for $d\geq 5$, and higher $N$ requires exponentially large computational resources.

\section{Trends of  GGM for multiqudit WGS }
\label{app:ggm}
The generalized geometric measure (GGM) \cite{GGM_wei2003,ASD_GGM2010,mixedggm2016}  quantifies the genuine multipartite entanglement content of a state. GGM of a pure state $\ket{\psi}$, denoted as $\mathcal{G(\ket{\psi})}$, is defined as
\begin{equation}
    \mathcal{G}(|\psi\rangle) = 1-\max_{A}\{\lambda^2_{A:V-A}\},
\end{equation}
where $\lambda_{A:V-A}$ is the maximum value of the Schmidt coefficient corresponding to $A:V-A$ bipartition. The maximization is performed over all possible bipartitions $A$ and $V-A$ of the whole system. As the number of qudits in a system grows, computing \( \mathcal{G} \) becomes increasingly challenging due to the growing number of partitions. However, an extensive numerical search indicates that if the system size is large enough, i.e.,, $N\geq 9$ for $d=3$, the maximum Schmidt coefficient that contributes in GGM is solely from the bipartition $A=\{1\}, B=\{2,3,4,\dots,N\}$ in $1$D with OBC. Similar results were previously obtained for $d=2$ in \cite{ghosh2023entanglement}.

\begin{figure*}
\centering
    \vspace{-3.0mm}
    \includegraphics[width=\linewidth]{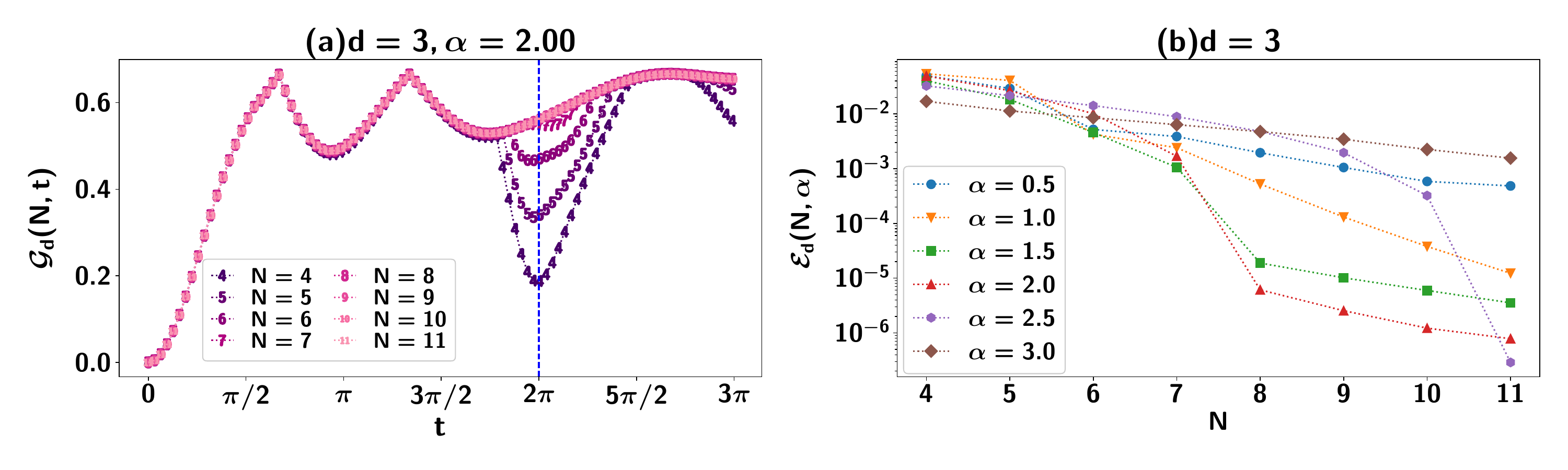}
    \vspace{-10.0mm}
    \caption{(a) GGM (ordinate) as a function of time (abscissa) for multiqutrit WGS. The exact value of GGM, $\mathcal{G}_d(N,t)$ is computed for $d=3$ and different system sizes at fixed $\alpha=2.0$. For most of the time until $t=2\pi$, GGM is almost insensitive to the system size, although GGM clearly depends on $N$ around $t=2\pi$. However, the dynamical GGM-profile  saturates for large system sizes even at  $t=2\pi$. (b) The average error, $\mathcal{E}_d(N, \alpha)$ (ordinate in logarithmic scale) between the actual GGM $\mathcal{G}_d$ and the approximate GGM $\bar{\mathcal{G}}_d$ for $d=3$, with system size $N$ (abscissa) of the WGS. The error decreases exponentially for large $N$ for small to moderate fall-off rates, $\alpha$. All the axes are dimensionless.}
    \label{fig:ggm_smallN}
\end{figure*}
\subsection{Numerical error analysis of GGM}
For $N$-qutrit ($d=3$)  WGS in a $1$D lattice with open ends, GGM is highly sensitive to the system size especially at small $N$ and near $t=2\pi$, as illustrated in Fig.~\ref{fig:ggm_smallN}. Such an $N$-dependence of GGM around $t=2\pi$ is seen for small to moderate values of  $\alpha$.

However, we argue that for sufficiently large \( N \), the error in computing the GGM by considering the maximum eigenvalue of \( \rho_1 \) is negligible. To compute the average error, we consider $\mathcal{E}_d(N, \alpha) = \langle|\mathcal{G}_d(N,\alpha, t)-\bar{\mathcal{G}}_d(N,\alpha, t)|\rangle_t$ where \( \mathcal{G}_d(N,\alpha,t) \) represents the exact value of the GGM, obtained by considering all possible bipartitions, and \( \bar{\mathcal{G}}_d(N,\alpha,t) \) denotes the GGM computed using only the eigenvalue of \( \rho_1 \). Although the time average in \( \mathcal{E}_d(N, \alpha) \) is taken over the interval \( t = 0 \) to \( t = 3\pi \), the qualitative behavior remains invariant even if we change the time interval. For $d=3$, the error $\mathcal{E}_d(N, \alpha)$ decreases exponentially with the system size $N$, as shown in Fig. \ref{fig:ggm_smallN}(b) for different values of $\alpha$. This suggests that GGM for a large system size $N$ can be computed efficiently from the single-site reduced density matrix of the edge for any $d$. In the case of \( d = 4 \), extensive numerical verification becomes challenging, and computations are restricted to \( N \leq 7 \); however, similar behavior was observed. Therefore, this analysis can be generalized for higher value of $d$ as well, and we safely assume $\mathcal{G}_d(N,\alpha,t)\equiv\bar{\mathcal{G}}_d(N,\alpha,t)$  for large system sizes, and can be computed from the single site density matrix given in Eq.~\eqref{eq:rho1}.

\begin{figure*}
\centering
    \includegraphics[width=\linewidth]{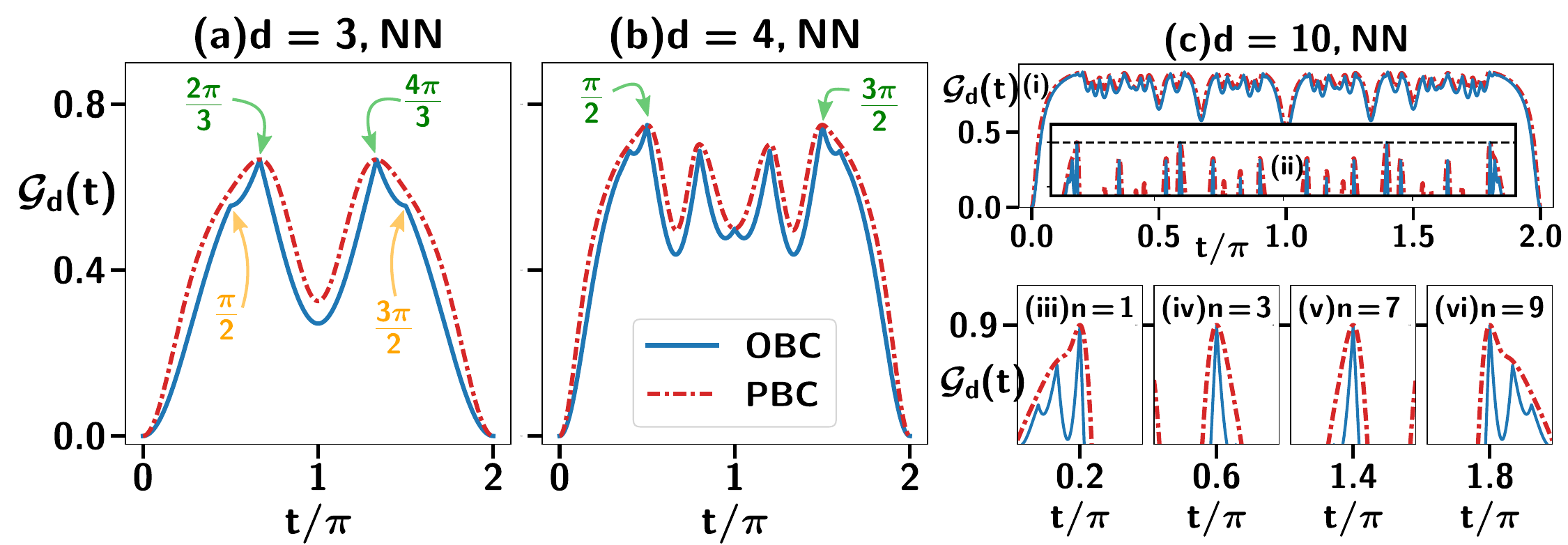}
    \vspace{-6.0mm}
    \caption{Profile of GGM, $\mathcal{G}_d$ (ordinate) with time $t$ in OBC [$\mathcal{G}^o_d(t)$] and PBC [$\mathcal{G}^p_d(t)$] with the nearest-neighbor interactions, showing nonanalytic behavior in OBC. (a) For $d=3$, while $\mathcal{G}^o_4(t)$ is non-analytic at various times $t$, including $t=\frac{2\pi}{3}, \frac{4\pi}{3}\equiv\frac{2\pi n}{3}$ (marked by green arrows), $\mathcal{G}^o_3(t)=\mathcal{G}^p_3(t)=\mathcal{G}^{\max}_3$. Other non-analytic $\mathcal{G}^o_3(t)$ occurs at $t=\frac{\pi}{2}, \frac{3\pi}{2}\equiv(m+\frac{1}{2})\pi$ but $\mathcal{G}^o_3(t)\neq \mathcal{G}^{\max}_3$ (marked by yellow arrows). (b) For $d=4$, while $\mathcal{G}^o_4(t)$ is non-analytic at various times $t$, including $t=\frac{\pi}{2}, \pi, \frac{3\pi}{2}\equiv\frac{2\pi n}{4}$, $\mathcal{G}^o_4(t)=\mathcal{G}^p_4(t)=\mathcal{G}^{\max}_d$ (marked with green arrows) only at $t=\frac{\pi}{2},\frac{3\pi}{2}=\frac{2\pi n}{d}$ ($n\in \mathbb{N}$) with $n$ and $d$ coprime. (c) The difference between a coprime ($n$,$d$) pair and non coprime pairs for $d=10$. (i) GGM profile with time for $d=10$, (ii) location of peaks in $\mathcal{G}_d(t)$, with dashed line showing $\mathcal{G}^{\max}_{10}=0.9$. These peaks are seen at $t_n=2\pi n/10$ for $n=1,3,7$ and $n=9$, as shown in (c)(iii)--(vi), illustrating non-analytic $\mathcal{G}^o_d$ and analytic $\mathcal{G}^p_d$ with $\mathcal{G}^o_d(t)=\mathcal{G}^p_d(t)=\mathcal{G}^{\max}_d$ at $t_n=2\pi n/d$, when $n$ and $d$ are coprimes. All the axes are dimensionless.}
    \label{fig:ggm_nn}
\end{figure*}

\subsection{Features of GGM for NN interaction}
We now show the profile of GGM, $\mathcal{G}(t)$, for the NN interactions ($\alpha\to\infty$), in an $1$D open lattice, to support the claims given in the main text. Using Eq.~\eqref{eq:rho1}, the single-site density matrices in NN interactions only, are given as 
\begin{eqnarray}
    [\rho^o_{\{1\}}(t)]_{\mu\nu}&=&\frac{1}{d^2}\sum_{k=0}^{d-1}\exp(i(\mu-\nu)kt), \label{eq:rho1obc}\\\relax
    [\rho^p_{\{1\}}(t)]_{\mu\nu}&=&\frac{1}{d^3}\bigg(\sum_{\mathclap{k=0}}^{d-1}\exp(i(\mu-\nu)kt)\bigg)^2\label{eq:rho1pbc}, 
\end{eqnarray}
where $\rho^o_{\{1\}}$ and $\rho^p_{\{1\}}$ are the density matrix of the first site with OBC and PBC respectively. For $d=2$, the corresponding eigenvalues can be computed analytically, giving $\mathcal{G}^{o}_2(t)=\frac{1}{2}\big(1-\big|\cos\frac{t}{2}\big|\big)$, and $\mathcal{G}^{p}_2(t)=\frac{1}{2}\big(1-\cos^2\frac{t}{2}\big)$. In OBC, the first site is at the edge, and gives the maximum Schmidt coefficient for GGM calculation, while for PBC all sites are equivalent due to translational symmetry. Note that in both cases, $\rho^{o(p)}_{\{1\}}$ have a time period of $2\pi$ in the case of NN interactions, and at $t=2n\pi$ ($n\in\mathbb{N}$), $\rho^{o(p)}_{\{1\}}=\ket{+}_d\bra{+}$,  since $(\mu-\nu), k\in \mathbb{Z}$, giving GGM $\mathcal{G}^{o(p)}_d(2n\pi)=0$. Interestingly, at $t=2\pi n/d$ ($n\in\mathbb{N}$), the summation in Eqs.~\eqref{eq:rho1obc} and \eqref{eq:rho1pbc} corresponds to the sum of the $d$th roots of unity only when $n$ and $d$ are coprimes along with $\mu\neq\nu$. Therefore, at $t=2\pi n/d$ when $n$ and $d$ are coprimes, $\rho^{o,p}_{\{1\}}=\mathbb{I}_d/d$  and $\mathcal{G}^{o,p}_d(2\pi n/d)=1-\frac{1}{d}\equiv\mathcal{G}^{\max}_d$, when $n$ and $d$ are coprimes. When \( n \) and \( d \) are not coprimes, there exist integers \( 0 \leq \mu \neq \nu < d \) such that \( (\mu - \nu) n / d \in \mathbb{Z} \). For this pair \( (\mu, \nu) \), the relation \( \exp(i (\mu - \nu) 2\pi n k/ d) = 1 \) holds and consequently,  \( \rho^o_{\{1\}}(2\pi n/d) = \rho^p_{\{1\}}(2\pi n/d) \neq \mathbb{I}_d \), which implies that \( \mathcal{G}^o(2\pi n/d) = \mathcal{G}^p(2\pi n/d) \neq \mathcal{G}^{\max}_d \) when \( n \) and \( d \) are not coprimes. The numerical behaviors of \( \mathcal{G}^o \) and \( \mathcal{G}^p \) are illustrated in Fig. \ref{fig:ggm_nn} for \( d = 3, 4, \) and \( 10 \) which confirms that \( \mathcal{G}^o_d(t) = \mathcal{G}^p_d(t) = \mathcal{G}^{\max}_d \) at \( t_n = \frac{2\pi n}{d} \), provided that \( n \) and \( d \) are coprimes.

\section{Analysis of GGM in two-dimensional lattice geometry} 
\label{app:ggm_2d}
Similar to $1$D with PBC, in $2$D too, all the sites being equivalent in terms of eigenvalues, the contribution to GGM comes again from a single party reduced density matrix. In contrast, under OBC, the contribution originates from any one of the four equivalent corner sites.

In $2$D square and Lieb lattice \cite{Guo2024}, we see that $\mathcal{G}_d^o(t)$ at $t=2\pi/d$, showing $\mathcal{G}_d^o(t_n)=\mathcal{G}_d^{\max}$ as a necessary and sufficient condition for the formation of a qudit cluster state with the NN interactions $(\alpha\to\infty)$.

In the presence of LR interactions, i.e., with finite $\alpha$ on $2$D lattices, similar results to $1$D are obtained, with $\delta\mathcal{G}_d^o=0$ when $t_{n,|\vec{r}|}=2\pi n|\vec{r}|^\alpha/d$ ($n\in \mathbb{N}$), ($n,d$) coprimes, with the possible values of $|\vec{r}|$ given by the lattice structure. For some $|\vec{r}|>1$, non-analytic $\mathcal{G}_d^o$ is also seen, e.g., at $t=2\pi, n=1$ with $d=2$, the square lattice has $|\vec{r}|=\sqrt2$ for the next-nearest neighbor, non-analyticity can be observed for $\alpha=\log_{\sqrt2}2=2$, while for the honeycomb lattice with $|\vec{r}|=\sqrt3$, it can be observed at $\alpha=\log_{\sqrt3}2\sim1.262$ \cite{ghosh2023entanglement}. Unlike the $1$D case, \( \mathcal{G}^o_d(t)\) achieves maximum value, \( \mathcal{G}^o_d(t) = \mathcal{G}^{\max}_d \), without any non-analyticity for certain $\vec{r}$, e.g., $\vec{r}=2$ in the $2$D square lattice, verifying that the multipartite entanglement profile can provide insight into information spreading during the evolution. 

Recently, multipartite entanglement has been shown to be a necessary condition for symmetry-protected topological order \cite{Lessa2025}. Our results also shed light from the non-analytic behavior of GGM in the nearest-neighbor case, while further study is required to confirm if the WGS in $2$D systems with LR interactions has any topological order and if its GGM can sufficiently detect any topological order.

\bibliography{bibf.bib}
\end{document}